\documentclass[amsmath, pra, twocolumn, showpacs]{revtex4-1}
\usepackage{graphicx}
\usepackage{dsfont}

\begin{document}     

\title{Propagation of nanofiber-guided light through an array of atoms}
 
\author{Fam Le Kien}

\author{A. Rauschenbeutel} 

\affiliation{Vienna Center for Quantum Science and Technology, Institute of Atomic and Subatomic Physics, Vienna University of Technology, Stadionallee 2, 1020 Vienna, Austria}

\date{\today}

\begin{abstract}
We study the propagation of nanofiber-guided light through an array of atomic cesium, taking into account the transitions between the hyperfine levels $6S_{1/2}F=4$ and $6P_{3/2}F'=5$ of the $D_2$ line. We derive the coupled-mode propagation equation, the input-output equation, the scattering matrix, the transfer matrix, and the reflection and transmission coefficients for the guided field in the linear, quasistationary, weak-excitation regime. We show that, when the initial distribution of populations of atomic ground-state sublevels is independent of the magnetic quantum number, the quasilinear polarizations along the principal axes $x$ and $y$, which are parallel and perpendicular, respectively, to the radial direction of the atomic position, are not coupled to each other in the linear coherent scattering process. 
When the guided probe field is quasilinearly polarized along the major principal axis $x$, forward and backward scattering have different characteristics. 
We find that, when the array period is far from the Bragg resonance, the backward scattering is weak. Under the Bragg resonance, most of the guided probe light can be reflected back in a broad region of field detunings even though there is an irreversible decay channel into radiation modes. When the atom number is large enough, two different band gaps may be formed, whose properties depend on the polarization of the guided probe field.
\end{abstract}

\pacs{42.50.Nn, 42.50.Ct, 42.81.Dp, 42.81.Gs}
\maketitle

\section{Introduction}
\label{sec:introduction}

Nanofibers are optical fibers that are tapered to a diameter comparable to or smaller than the wavelength of light \cite{Mazur's Nature,Birks,taper}. In such a thin fiber, the guided field penetrates an appreciable distance into the surrounding medium in the form of an evanescent wave carrying a significant fraction of the propagation power and having a complex polarization pattern \cite{Bures99,Tong04,fibermode}. Nanofibers have attracted considerable attention for a wide range of potential practical applications \cite{Morrissey13}.
Nanofiber-guided light fields find applications for trapping atoms \cite{fiber trap,Vetsch10,Goban12}, for probing atoms \cite{Domokos02,absorption,Fam14,Nayak07,Nayak09,Dawkins11,Reitz13,Russell13,Reitz14,Mitsch14a,Mitsch14b}, molecules \cite{Stiebeiner09}, quantum dots \cite{Yalla12}, and color centers in nanodiamonds \cite{Schroder12,Liebermeister13}, and for mechanical manipulations of small particles \cite{Skelton12,Brambilla07,Fam13,Chormaic14}. 

Various applications in both fundamental and applied physics requires the ability to control and manipulate atoms \cite{Schlosser,Kuhr,Sackett}. In order to find an effective way to work with atoms trapped outside a nanofiber, we need to know the optical response of the atoms to a near-resonant guided field propagating along the fiber. The absorption and scattering of guided light by a single atom have been studied \cite{Domokos02,absorption,Fam14}. It has been shown for a two-level atom \cite{Domokos02} and a multilevel cesium atom \cite{absorption} that, when the transverse extension of the probe field in a guided mode is close to the radiative cross section of the atom, the latter becomes a significant scatterer. 

Recent experimental progress has demonstrated that the scattering of guided light from  realistic atoms is very different from the case of atoms in free space \cite{Reitz14,Mitsch14a,Mitsch14b}. It has been shown that, due to  the existence of a longitudinal component of the guided-mode profile function and
the complex transition structure of a realistic atom, the rate of scattering of guided light from the atom into the guided modes 
is asymmetric with respect to the forward and backward directions and depends on the polarization of the probe field \cite{Fam14}. Also, it has recently been demonstrated experimentally that spin-orbit coupling of guided light can lead to directional spontaneous emission \cite{Mitsch14b}.

The theory of propagation of guided light with complex polarization in an atomic medium around a nanofiber has been developed \cite{Fam09}. In that theory, the guided probe field interacts with a single atomic transition and the medium is disordered. Meanwhile, the experiments with atom-waveguide interfaces \cite{Dawkins11,Reitz13,Reitz14,Mitsch14a,Mitsch14b} used linear arrays of atoms prepared in a nanofiber-based optical dipole trap \cite{Vetsch10,Goban12}. The theory of  \cite{Fam09} can be used for atomic arrays when the array period is far from the Bragg resonance condition.
However, when the array period is near to the Bragg resonance, the discreteness and regularity of the array may lead to  significant effects, such as nearly perfect atomic mirrors, photonic band gaps,
long-range interaction, and self-ordering \cite{Deutsch95,Birkl95,Henkel03,Artoni05,Petrosyan07,Schilke11,Schilke12,Chang07,Chang11,Chang12,Ritsch14a,Ritsch14b}.  
In the prior work for a periodic array of atoms along a waveguide  \cite{Chang11,Chang12}, two-level atoms and scalar guided light fields were considered. In view of the recent results and insights, it is necessary to develop a systematic theory for the propagation of guided light in an atomic array taking into account the vector nature of the guided field, the multilevel structure of the atoms, and the discreteness and periodicity of the array. 

In this paper, we study the propagation of nanofiber-guided light through an array of multilevel atomic cesium. We derive the coupled-mode propagation equation, the input-output equation, the scattering matrix, the transfer matrix, and the reflection and transmission coefficients
for the guided field in the linear, quasistationary, weak-excitation regime. In our treatment, we take into account the specific polarization of the guided field, the multilevel structure of the atoms, and the discreteness and periodicity of the atomic positions in the array.  
   
The paper is organized as follows. 
In Sec.\ \ref{sec:theory} we present a general theory for the propagation of guided light in a linear atomic array. 
In Sec.\ \ref{sec:quasilinear} we investigate the reflection and transmission of quasilinearly polarized guided light. 
In Sec.\ \ref{sec:numerical} we present the results of numerical calculations. 
Our conclusions are given in Sec.~\ref{sec:summary}.

\section{Propagation of guided light in a linear array of atoms}
\label{sec:theory}

\begin{figure}[tbh]
\begin{center}
  \includegraphics{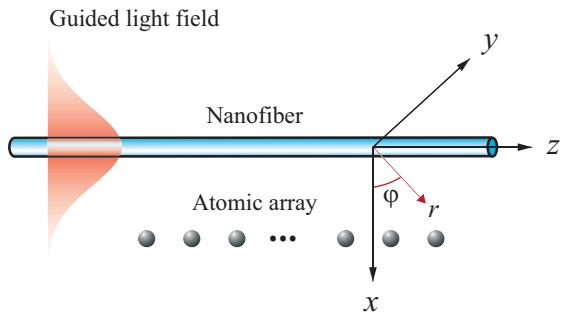}
 \end{center}
\caption{(Color online) Probing an array of atoms by a guided light field propagating along a thin optical fiber.}
\label{fig1}
\end{figure}

We consider the propagation of a guided light field through a linear array of alkali-metal atoms trapped along a nanofiber (see Fig.~\ref{fig1}). The thin cylindrical silica fiber with radius $a$ and refractive index $n_1$ is surrounded by vacuum with refractive index $n_2=1$. Although our theory is general and applicable, in principle, to arbitrary multilevel atoms, we assume cesium atoms throughout this paper. For simplicity, 
we neglect the effect of the surface-induced potential on the atomic energy levels. This approximation is reasonable when the atoms are not close to the fiber surface \cite{Fam07}.

%%%%%%%%%%%%%%%%%%%%%%% Figure 1

\subsection{Interaction of a single atom with guided light}
\label{subsec:interaction}

We use the Cartesian coordinates  $\{x,y,z\}$ and the associated cylindrical coordinates $\{r,\varphi,z\}$, with $z$ being the fiber axis (see Fig.~\ref{fig1}). We represent the electric component of the guided light field as
$\mathbf{E}=(\boldsymbol{\mathcal{E}}e^{-i\omega_L t}+\mathrm{c.c.})/2
=(\mathcal{E}\mathbf{u}e^{-i\omega_L t}+\mathrm{c.c.})/2$, where $\omega_L$ is the angular frequency and $\boldsymbol{\mathcal{E}}=\mathcal{E}\mathbf{u}$ is the slowly varying envelope of the positive-frequency part, with $\mathcal{E}$ and $\mathbf{u}$ being the field amplitude and the polarization vector, respectively. We assume that the guided probe field $\mathbf{E}$ propagates in the positive direction $+z$, from the left-hand side to the right-hand side of Fig.~\ref{fig1}.
In general, the amplitude $\mathcal{E}$ is a complex scalar and the polarization vector $\mathbf{u}$ is a complex unit vector. 

We study the $D_2$ line of atomic cesium, which  occurs at the wavelength $\lambda_0=852$ nm and 
corresponds to the transition from the ground state $6S_{1/2}$
to the excited state $6P_{3/2}$. We assume that the cesium atoms are initially prepared in the hyperfine-structure (hfs) level $F=4$ of the ground state $6S_{1/2}$ and that the probe field is tuned close to resonance with the transition 
from this ground-state hfs level to the hfs level $F'=5$ of the excited state  $6P_{3/2}$.
Among the hfs components  of the $D_2$ line,  
the transition $6S_{1/2}F=4\leftrightarrow 6P_{3/2}F'=5$ has the strongest oscillator strength.
Because of the selection rule
$\Delta F=0,\pm1$, spontaneous emission from the excited hfs level $6P_{3/2}F'=5$ to the ground state is always to the hfs level $6S_{1/2}F=4$, not
to the other hfs level $6S_{1/2}F=3$. Therefore, the magnetic sublevels of the hfs levels $6S_{1/2}F=4$ and $6P_{3/2}F'=5$ form a closed set, which is used for laser cooling in magneto-optical traps \cite{coolingbook}.

In order to describe the internal state of the cesium atoms, we use the fiber axis $z$ as the quantization axis.
In addition, we assume that the atoms are located on the positive side of the axis $x$. 
For convenience, we introduce the notations 
$|e\rangle\equiv|F'M'\rangle$ and $|g\rangle\equiv|FM\rangle$ for the Zeeman (magnetic) sublevels 
$F'M'$ and $FM$ of the hfs levels $6P_{3/2}F'=5 $ and $6S_{1/2}F=4$, 
respectively.  
The spherical tensor components of the  dipole matrix-element vector for the transition between 
$|F'M'\rangle$ and $|FM\rangle$ are given by \cite{Shore}
\begin{eqnarray}\label{x3}
d_{M'M}^{(q)}&=&(-1)^{I+J'-M'}\langle J' \| D\| J\rangle\sqrt{(2F+1)(2F'+1)}\nonumber\\
&&\mbox{}\times
\begin{Bmatrix}J'&F'&I\\F&J&1\end{Bmatrix}
\begin{pmatrix}F&1&F'\\M&q&-M'\end{pmatrix},
\end{eqnarray}
where $q=M'-M=0,\pm1$. In Eq.~\eqref{x3}, the array in the curly braces is a 6$j$ symbol, 
the array in the parentheses is a 3$j$ symbol, 
and $\langle J' \| D\| J\rangle$ is the reduced electric-dipole matrix element in the $J$ basis.
For the cesium $D_2$ line, we have $\langle J' \| D\| J\rangle=6.347$ a.u. 
$=5.38\times10^{-29}$ C m \cite{coolingbook}.

We introduce the notation $\mathcal{E}_{q}$ with $q=0,\pm1$ for the spherical tensor components of the field envelope vector $\boldsymbol{\mathcal{E}}$,
that is, $\mathcal{E}_{-1}=(\mathcal{E}_x-i\mathcal{E}_y)/\sqrt{2}$, $\mathcal{E}_0=\mathcal{E}_z$, and $\mathcal{E}_{1}=-(\mathcal{E}_x+i\mathcal{E}_y)/\sqrt{2}$.
We assume that the probe field is a classical coherent laser field.
The interaction of a single atom with the probe field is characterized by the set of Rabi frequencies 
\begin{equation}\label{x4}
\Omega_{eg}=\frac{1}{\hbar}(\mathbf{d}_{eg}\cdot\boldsymbol{\mathcal{E}})
=\frac{1}{\hbar}\sum_{q=0,\pm1} (-1)^q d_{eg}^{(q)}\mathcal{E}_{-q}.
\end{equation}
The time evolution of the reduced density operator $\rho$ of the atom
is governed by the equations \cite{absorption}
\begin{eqnarray}\label{x5}
\dot{\rho}_{ee'}&=&\frac{i}{2}\sum_{g}(\Omega_{eg}\rho_{ge'}
-\Omega_{e'g}^*\rho_{eg})\nonumber\\&&\mbox{}
-\frac{1}{2}\sum_{e''}(\gamma^{(\mathrm{tot})}_{ee''}{\rho}_{e''e'}+\gamma^{(\mathrm{tot})}_{e''e'}{\rho}_{ee''}),\nonumber\\
\dot{\rho}_{gg'}&=&-\frac{i}{2}\sum_{e}(\Omega_{eg'}
\rho_{ge}-\Omega_{eg}^*\rho_{eg'})
+\sum_{ee'}\gamma^{(\mathrm{tot})}_{e'eg'g}{\rho}_{ee'},\nonumber\\
\dot{\rho}_{eg}&=&i\delta\rho_{eg}
+\frac{i}{2}\sum_{g'}\Omega_{eg'}\rho_{g'g}
-\frac{i}{2}\sum_{e'}\Omega_{e'g}\rho_{ee'}
\nonumber\\&&\mbox{} 
-\frac{1}{2}\sum_{e'} \gamma^{(\mathrm{tot})}_{ee'}\rho_{e'g}.
\end{eqnarray}
Here,  $\delta=\omega_L-\omega_0$ is the detuning of the field from the atomic transition frequency $\omega_0=\omega_e-\omega_g$. 
The coefficients $\gamma^{(\mathrm{tot})}_{ee'gg'}$ and $\gamma^{(\mathrm{tot})}_{ee'}$ characterize the effect of spontaneous emission on the reduced density operator of the atomic state. They are given as  \cite{cesium decay}
$\gamma^{(\mathrm{tot})}_{ee'gg'}=\gamma^{(\mathrm{gyd})}_{ee'gg'}+\gamma^{(\mathrm{rad})}_{ee'gg'}$
and $\gamma^{(\mathrm{tot})}_{ee'}=\sum_{g}\gamma^{(\mathrm{tot})}_{ee'gg}=\gamma^{(\mathrm{gyd})}_{ee'}+\gamma^{(\mathrm{rad})}_{ee'}$.
Here,  the set of coefficients $\gamma^{(\mathrm{gyd})}_{ee'gg'}$ and $\gamma^{(\mathrm{gyd})}_{ee'}=\sum_{g}\gamma^{(\mathrm{gyd})}_{ee'gg}$ describes spontaneous emission into guided modes, and the set of coefficients $\gamma^{(\mathrm{rad})}_{ee'gg'}$ and 
$\gamma^{(\mathrm{rad})}_{ee'}=\sum_{g}\gamma^{(\mathrm{rad})}_{ee'gg}$ describes spontaneous emission into radiation modes. 
The total decay rate of the population of the excited magnetic sublevel $|e\rangle$ is 
$\gamma^{(\mathrm{tot})}_{ee}=\gamma^{(\mathrm{gyd})}_{ee}+\gamma^{(\mathrm{rad})}_{ee}$.
The explicit expressions for the decay coefficients are given in Ref.~\cite{cesium decay} and are summarized in Appendixes \ref{sec:guided} and  \ref{sec:radiation}.
We assume that the atoms are initially prepared in an incoherent mixture of the magnetic sublevels 
$|M\rangle$ of the ground-state hyperfine level $F$ and that the initial population distribution of the atoms is flat with respect to $M$. 
We are interested in the regime where the probe field $\boldsymbol{\mathcal{E}}$ is stationary and the atoms are weakly excited.

\subsection{Photon flux amplitude}
\label{subsec:photon flux}

We now consider the time evolution of guided light interacting with a linear array of atoms.
We treat the field quantum mechanically in this subsection. 
However, in the next subsection, we will replace the quantum field by a classical field. 
The use of the quantum description is convenient for deriving the coupled-mode propagation equation for the field.

We first consider a single mode of the guided light field with the propagation direction $f$ and 
the polarization $p$.
The index $f=+1$ or $-1$ (or simply $+$ or $-$) stands for the forward ($+\hat{\mathbf{z}}$) or backward ($-\hat{\mathbf{z}}$) propagation direction, respectively.
The index $p$ is $p=l$ or $p=\xi$ for the quasicircular or quasilinear polarization, respectively. 
The index $l=+1$ or $-1$ (or simply $+$ or $-$) refers respectively to the counterclockwise or
clockwise circulation of the transverse component of the field with respect to the positive direction of the fiber axis $z$.
The index $\xi=x$ or $y$ refers to the so-called major or minor principal axis, respectively, 
which is parallel or perpendicular, respectively, to the radial direction of the positions of the atoms in the array. 

We assume that the spread in frequency of the field around its central frequency $\omega_L$ is sufficiently small that the fiber dispersion can be neglected. The flux of energy (power) of the field propagating along the fiber axis $z$ is given by 
$P_z=\int\mathbf{S}\cdot \hat{\mathbf{z}} \; d^2\mathbf{r}$. Here,  $\mathbf{S}$ is the Poynting vector and $\int d^2\mathbf{r}=\int_0^{2\pi}d\varphi\int_0^{\infty}r\,dr$ is the integral over the fiber cross-section plane. In the framework of the rigorous fiber theory \cite{fiber books} and the continuous-mode field quantization  \cite{continuous,Domokos02,Loudon}, the quantum-mechanical expression for the energy flux $P_z$ is given by  \cite{fibercorr}
$P_z(z,t)=\hbar\omega_L A_{fp}^\dagger(z,t) A_{fp}(z,t)$,
where 
\begin{equation}\label{x14}
A_{fp}(z,t)=\frac{1}{\sqrt{2\pi}}\int_0^{\infty}d\omega\, a_{\omega fp}(t) e^{if\beta(\omega) z}
\end{equation}
is the generalized Fourier transform of the photon operator $a_{\omega fp}(t)$ in the continuous-mode field quantization \cite{continuous,Domokos02,Loudon}. Here,  $\beta=\beta(\omega)$ is the longitudinal propagation constant of the guided field and is a function of the mode frequency $\omega$ \cite{fiber books}. 
The photon operator $A_{fp}(z,t)$ describes the annihilation of a guided photon in the time-space domain \cite{continuous,Domokos02,Loudon,fibercorr}. This operator characterizes the amplitude of the energy flux of the guided field. 
The photon operators $a_{\omega fp}(t)$ and $a_{\omega fp}^\dagger(t)$ are the time-dependent operators in the Heisenberg picture. They satisfy the continuous-mode bosonic commutation rules
$[a_{\omega fp}(t),a_{\omega' f'p'}^\dagger(t)]=\delta(\omega-\omega')\delta_{ff'}\delta_{pp'}$. 

We label the atoms by the index $j$. Each atom $j$ has a set of upper levels $|e_j\rangle$ with energy $\hbar\omega_{ej}=\hbar\omega_{e}$ and a set of lower levels $|g_j\rangle$  with energy $\hbar\omega_{gj}=\hbar\omega_{g}$. We introduce the notation $\sigma_{abj}=|a_j\rangle\langle b_j|$, where $a_j$ and $b_j$ can be $e_j$ or $g_j$.
The downward and upward transitions of the atoms are described by the operators $\sigma_{gej}=|g_j\rangle\langle e_j|$ 
and $\sigma_{egj}=\sigma_{gej}^\dagger=|e_j\rangle\langle g_j|$, respectively. 
The atoms are located at the positions with the Cartesian coordinates $(x_j,y_j,z_j)$ or the corresponding cylindrical coordinates $(r_j,\varphi_j,z_j)$. We assume that the spatial displacements of the atoms during the interaction time can be neglected.

The Hamiltonian for the interaction between the atoms and the guided light field in the dipole and rotating-wave approximations is given by 
\begin{equation}\label{x15}
H_{\mathrm{int}}=-\frac{i\hbar}{\sqrt{2\pi}}\int _0^{\infty}d\omega \sum_{fpegj} \mathcal{G}_{\omega fp egj}
\sigma_{egj} a_{\omega fp}e^{if\beta z_j}+\mbox{H.c.},
\end{equation}
where 
\begin{equation}\label{x16}
\mathcal{G}_{\omega fl egj}=\sqrt{\frac{\omega}{2\epsilon_0\hbar v_g}}\;
\big[\mathbf{d}_{egj}\cdot\mathbf{e}^{(\omega fl)}(r_j,\varphi_j)\big]e^{il\varphi_j}
\end{equation}  
are the coupling coefficients for quasicircularly polarized guided modes with the polarization index $l=\pm$ and 
\begin{equation}\label{x17}
\mathcal{G}_{\omega f\xi egj}=\sqrt{\frac{\omega}{2\epsilon_0\hbar v_g}}\;
\big[\mathbf{d}_{egj}\cdot\mathbf{e}^{(\omega f\xi)}(r_j,\varphi_j)\big]
\end{equation}
are the coupling coefficients for quasilinearly polarized guided modes with the polarization index $\xi=x,y$.  
Here,  $\mathbf{e}^{(\omega fl)}$ and $\mathbf{e}^{(\omega f\xi)}$ are the mode profile functions, 
$v_g=d\omega/d\beta$ is the group velocity of the guided field, and
$\mathbf{d}_{egj}$ is the matrix element of the electric dipole vector for the transition $|e_j\rangle\leftrightarrow|g_j\rangle$ of atom $j$. 
The explicit expressions for the guided-mode profile functions $\mathbf{e}^{(\omega fl)}$ and $\mathbf{e}^{(\omega f\xi)}$
are given in Refs.~\cite{fiber books,fibermode,cesium decay} and are summarized in Appendix \ref{sec:guided}. 

It has been shown in Refs.~\cite{fibercorr,Domokos02} that, under the condition of weak interaction and narrow bandwidth, the explicit expression for the energy flux amplitude $A_{fp}$ is
\begin{eqnarray}\label{x18}
A_{fp}(z,t)&=&A_{fp}^{(\mathrm{in})}(z,t)
+\sum_{egj} \mathcal{G}_{fpegj}^* 
\sigma_{gej}(t-|z-z_j|/v_g)
\nonumber\\&&\mbox{}\times
\exp[i\omega_L(1/v_p -1/v_g)|z-z_j|]
\nonumber\\&&\mbox{}\times
\Theta[f\;(z-z_j)]\Theta(t-t_0-|z-z_j|/v_g).
\end{eqnarray}
Here,  
\begin{equation}\label{x19}
A_{fp}^{(\mathrm{in})}(z,t)=\frac{1}{\sqrt{2\pi}}\int_0^{\infty}d\omega\, a_{\omega fp}(t_0)e^{-i\omega(t-t_0)} e^{if\beta z}
\end{equation}
is the injected field, with $t_0$ being the initial time, 
$\Theta(x)$ stands for the Heaviside step function, equal to zero for negative argument and one for positive argument, and $v_p=\omega/\beta$ is the phase velocity.
The group velocity $v_g$ and the phase velocity $v_p$ are evaluated at the central frequency $\omega_L$ of the probe field.
The notation $\mathcal{G}_{fpegj}\equiv\mathcal{G}_{\omega_L fpegj}$ stands for the coupling coefficients estimated at the central frequency $\omega_L$ of the field.
When the detuning of the probe field from the atomic transition is small as compared to the atomic transition frequency $\omega_0$ and to the optical field frequency $\omega$, we have $\mathcal{G}_{fpegj}\simeq\mathcal{G}_{\omega_0 fpegj}$.

\subsection{Scattering matrix and coupled-mode propagation equation}
\label{subsec:propagation equation}

In order to derive the propagation equation for the coupled photon flux amplitudes $A_{fp}(z,t)$, we differentiate expression (\ref{x19}) with respect to $z$ and $t$ separately. When we combine the results, we find 
\begin{eqnarray}\label{x20}
\left[\frac{\partial}{f\partial z}
+i\omega_L(1/v_g-1/v_p)
+\frac{\partial}{v_g\partial t}\right] A_{fp}(z,t)
\nonumber\\
=\sum_{egj} \mathcal{G}_{fpegj}^* \sigma_{gej}(t)\delta(z-z_j).
\end{eqnarray}
We introduce the variables
\begin{eqnarray}\label{x21}
\tilde{A}_{fp}(z,t)&=& A_{fp}(z,t) e^{i\omega_Lt},
\nonumber\\
\tilde{\sigma}_{gej}(t)&=& \sigma_{gej}(t)e^{i\omega_Lt},
\nonumber\\
\tilde{\sigma}_{ee'j}(t)&=&\sigma_{ee'j}(t),
\nonumber\\
\tilde{\sigma}_{gg'j}(t)&=&\sigma_{gg'j}(t),
\end{eqnarray}
which vary slowly in time.
Then, we can transform Eq.~(\ref{x20}) to
\begin{eqnarray}\label{x22}
\lefteqn{\left(\frac{\partial}{f\partial z}-i\beta_L+\frac{\partial}{v_g\partial t}\right) \tilde{A}_{fp}(z,t)}
\nonumber\\&&
=\sum_{egj} \mathcal{G}_{fpegj}^* \tilde{\sigma}_{gej}(t)\delta(z-z_j).
\end{eqnarray}
We note that, unlike the conventional one-dimensional propagation equations, the expression on the left-hand side of Eq.~\eqref{x22} contains the terms 
$(-i\beta_L+v_g^{-1}\partial/\partial t)\tilde{A}_{fp}$. The reason is that the temporal optical modulations of the photon amplitude $A_{fp}$ are removed from the envelope $\tilde{A}_{fp}$ but the spatial optical modulations are still kept. It is convenient to keep the spatial optical dependence when we study the coupling of different modes 
with different propagation directions.

We introduce the atom number density
\begin{equation}\label{x23} 
n_A=n_A(r,\varphi,z)=\sum_j \delta(x-x_j)\delta(y-y_j)\delta(z-z_j).
\end{equation}
We replace an arbitrary discrete-variable function $\Phi_j$ by a continuous-variable function $\Phi(r,\varphi,z)$. The summation over the atomic label $j$ can be formally replaced
by the integration over the space with the help of the formula $\sum_j \Phi_j=\int_{-\infty}^\infty dz \int d^2\mathbf{r}\; n_A(r,\varphi,z) \Phi(r,\varphi,z)$. 
Then, Eq.~(\ref{x22})  becomes
\begin{equation}\label{x24}
\left(\frac{\partial}{f\partial z}-i\beta_L+\frac{\partial}{v_g\partial t}\right) \tilde{A}_{fp}=\int d^2\mathbf{r}\; n_A \sum_{eg} \mathcal{G}_{fpeg}^* \tilde{\sigma}_{ge}.
\end{equation}
We note that the derivative with respect to the axial coordinate $z$ and the integral over the fiber cross-section plane $\mathbf{r}=(x,y)$ in the above equation are formal
because the atom number density $n_A$, given by Eq.~\eqref{x23}, contains the Dirac delta function. 
The use of this generalized-function technique is a convenient way to describe the sum over the discrete atom label $j$. 

The mean value of the photon flux is given by $\mathcal{N}_{fp}=\langle \tilde{A}_{fp}^\dagger \tilde{A}_{fp}\rangle$.
We now assume that the guided probe light field is in a coherent state or is simply a classical field. 
In this case, we have $\mathcal{N}_{fp} = |\langle \tilde{A}_{fp}\rangle|^2$.
In order to get a propagation equation
for such a field, we average Eq.~(\ref{x24}) over the quantum state of the combined atom-field system. When we use the notation $\mathcal{A}_{fp}=\langle \tilde{A}_{fp}\rangle=\langle A_{fp}(z,t)\rangle e^{i\omega_Lt}$ for the mean value of the envelope of the photon flux amplitude and the notation 
$\rho_{eg}=\langle \tilde{\sigma}_{ge}\rangle=\langle \sigma_{ge}(r,\varphi,z,t)\rangle e^{i\omega_Lt}$ for the atomic transition coherence, we obtain
\begin{equation}\label{x25}
\left(\frac{\partial }{f\partial z}-i\beta_L+\frac{\partial}{v_g\partial t}\right)\mathcal{A}_{fp}=\int d^2\mathbf{r}\; n_A \sum_{eg}\mathcal{G}_{fpeg}^* \rho_{eg}.
\end{equation}
We now assume that the field envelope $\mathcal{A}_{fp}$ is a slowly varying function of time $t$. In this case, we can neglect the time derivative in Eq.~\eqref{x25} and obtain
\begin{equation}\label{x26}
\left(\frac{\partial }{f\partial z}-i\beta_L\right)\mathcal{A}_{fp}=\int d^2\mathbf{r}\; n_A \sum_{eg}\mathcal{G}_{fpeg}^* \rho_{eg}.
\end{equation}
We use the approximate propagation equation \eqref{x26} in what follows.

Equation \eqref{x26} is the axial (one-dimensional) propagation equation for 
the amplitude of the photon flux of the guided light field. It is interesting to note that, even though the field in a guided mode has three nonzero spherical tensor components and is an evanescent wave  outside the fiber surface, the propagation of the field through a gas medium can be described by a one-dimensional equation. This propagation equation incorporates the complexity of the polarization vector structure and the evanescent-wave nature of the mode profile.

It follows from the Hamiltonian (\ref{x15}) and the transformation (\ref{x21}) that the time evolution of the density matrix elements of the atom is governed by Eqs.~\eqref{x5}.
In terms of the photon flux amplitudes $\mathcal{A}_{fp}$, the Rabi frequencies $\Omega_{eg}$ are given as 
\begin{equation}\label{x27}
\Omega_{eg}=2i\sum_{fp}\mathcal{G}_{fpeg}\mathcal{A}_{fp}.
\end{equation}
 
We consider the quasistationary regime where the atomic transition coherence $\rho_{eg}$ adiabatically follows the driving field.
This regime occurs when the characteristic atomic decay rate $\gamma$ or the field detuning $\delta$ is large as compared to the characteristic Rabi frequency $\Omega$, and the interaction time $t$ is large as compared to the characteristic atomic lifetime $\tau=\gamma^{-1}$.
In this regime, we can neglect the time derivative of $\rho_{eg}$ in the last equation in Eqs.~\eqref{x5}. 
Then, we obtain
\begin{equation}\label{x28}
i\delta\rho_{eg}-\frac{1}{2}\sum_{e'} \gamma^{(\mathrm{tot})}_{ee'}\rho_{e'g}=
\frac{i}{2}\sum_{e'}\Omega_{e'g}\rho_{ee'}-\frac{i}{2}\sum_{g'}\Omega_{eg'}\rho_{g'g}.
\end{equation}
We assume that the initial state of the atom is an incoherent mixture of the Zeeman sublevels of a hyperfine level $F$ the ground state, that is,
$\rho_{gg'}(t=0)=\delta_{gg'}p_g$. Here,  $p_g\geq0$ and $\sum_g p_g=1$. In the particular case where the initial population distribution $p_g$ is flat, we have $p_g=1/(2F+1)$.
We assume that the interaction time is  small as compared to the characteristic Rabi period
$\Omega^{-1}$ so that the effects of optical pumping are weak and, consequently, the deviations of the atomic populations from the initial values are small.
In this case, we can use the approximations $\rho_{ee'}(t)\simeq0$ and $\rho_{gg'}(t)\simeq\rho_{gg'}(t=0)=\delta_{gg'}p_g$ to calculate the atomic transition coherence $\rho_{eg}$ from Eq.~\eqref{x28}. 
Furthermore, we assume that the atom is not too close to the fiber. In this case, the effect of the fiber on the spontaneous emission rate is weak and, hence, we can use the generalized free-space expression $\gamma^{(\mathrm{tot})}_{ee'}=\delta_{ee'}\gamma^{(\mathrm{tot})}$ for the decay coefficients of the multilevel atom. 
Here,  $\gamma^{(\mathrm{tot})}=\gamma^{(\mathrm{gyd})}+\gamma^{(\mathrm{rad})}$ is the sublevel-averaged fiber-enhanced spontaneous emission rate, which contains the contributions $\gamma^{(\mathrm{gyd})}$ and $\gamma^{(\mathrm{rad})}$
from the radiation and guided modes, respectively \cite{cesium decay}.
From now on, we drop the upper label of $\gamma^{(\mathrm{tot})}$, that is, we use the notation $\gamma=\gamma^{(\mathrm{tot})}=\gamma^{(\mathrm{gyd})}+\gamma^{(\mathrm{rad})}$.
Then, we find 
\begin{equation}\label{x29}
\rho_{eg}= -p_g\frac{\Omega_{eg}}{2\delta+i\gamma}= - \frac{2}{\gamma-2i\delta}p_g\sum_{fp}\mathcal{G}_{fpeg}\mathcal{A}_{fp}.
\end{equation}
We insert Eq.~(\ref{x29}) into Eq.~(\ref{x26}). Then, we obtain the coupled-mode propagation equation
\begin{equation}\label{x30}
\left(\frac{\partial }{\partial z}-if\beta_L\right)\mathcal{A}_{fp}= -\int d^2\mathbf{r}\; n_A  \sum_{f'p'} S_{fpf'p'}
\mathcal{A}_{f'p'},
\end{equation}
where
\begin{equation}\label{x31}
S_{fpf'p'}=\frac{2f}{\gamma-2i\delta}\sum_{eg} p_g \mathcal{G}_{fpeg}^*\mathcal{G}_{f'p'eg}
\end{equation}
characterizes scattering of guided light from the mode $f'p'$ into the mode $fp$ or vice versa by a single atom in the linear quasistationary weak-excitation regime.
We note that, since the emission of an atom into a guided mode is much weaker than the total emission into all the modes, 
we have $|\mathcal{G}_{fpeg}^*\mathcal{G}_{f'p'eg}|\ll\gamma$, which leads to $|S_{fpf'p'}|\ll1$.

The optical depth per atom of the linear atomic array interacting with a single guided mode $fp$ is given by
\begin{equation}\label{x32}
\mathcal{D}=2\mathrm{Re}(fS_{fpfp})= \frac{4\gamma}{\gamma^2+4\delta^2}\sum_{eg} p_g |\mathcal{G}_{fpeg}|^2.
\end{equation}
In general, the optical depth per atom $\mathcal{D}$ depends on the propagation direction $f=\pm$ and the polarization $p$ of the guided probe field. Due to the evanescent-wave profile of the guided field in the fiber transverse plane,  $\mathcal{D}$ decreases with increasing distance $r-a$ of the array from the fiber surface. 

We introduce the notation $\mathbf{S}$ for the square matrix consisting of the matrix elements $S_{fpf'p'}$ with the row index $fp$ and the column index $f'p'$, 
the notation $\boldsymbol{\mathcal{A}}$ for the vector 
consisting of the components $\mathcal{A}_{fp}$ with the index $fp$,
and the notation $\mathbf{B}$ for the diagonal square matrix consisting of the matrix elements 
$B_{fpf'p'}=\delta_{ff'}\delta_{pp'}f\beta_L$
with the row index $fp$ and the column index $f'p'$. 
Then, we can rewrite Eq.~(\ref{x30}) in the matrix form
\begin{equation}\label{x33}
\frac{\partial }{\partial z}\boldsymbol{\mathcal{A}}= \left(i\mathbf{B}-\int d^2\mathbf{r}\; n_A  \mathbf{S}\right)
\boldsymbol{\mathcal{A}}.
\end{equation}
The formal expression for the solution of Eq.~\eqref{x33} is
\begin{equation}\label{x34}
\boldsymbol{\mathcal{A}}(z)=\hat{\mathcal{P}}\exp\left(i\int_0^z \mathbf{k}_A(z') dz'\right)
\boldsymbol{\mathcal{A}}(0),
\end{equation}
where $\mathbf{k}_A=\mathbf{B}+i\int d^2\mathbf{r}\; n_A  \mathbf{S}$ is the propagator
and $\hat{\mathcal{P}}$ is the path-ordering operator with respect to the axial coordinate $z'$ \cite{path ordering}.

The matrix $\mathbf{S}$ is called the single-atom guided-field scattering matrix. 
In order to get insight into the properties of $\mathbf{S}$, we calculate the matrix elements $S_{f\xi f'\xi'}$ in the mode basis formed by quasilinearly polarized modes
with the indices $f=\pm$ and $\xi=x,y$. As shown in Appendix \ref{sec:properties}, 
when the atom is positioned on the $x$ axis and the initial population distribution $p_g$ is flat, we have
\begin{equation}\label{x35} 
S_{fxf'y}=S_{fyf'x}=0. 
\end{equation}
Thus, the quasilinear polarizations along the principal axes $x$ and $y$ (which are parallel and perpendicular 
to the radial direction of the atomic position, respectively) are not coupled to each other in the linear coherent scattering process.
Furthermore, according to Appendix \ref{sec:properties}, the nonzero matrix elements of the scattering matrix $\mathbf{S}$ are
\begin{subequations}\label{x36}
\begin{eqnarray}
S_{fxf'x}&=&\frac{fu_0}{\gamma-2i\delta}
\big(|e_r|^2+ff'|e_z|^2 \big),\label{x36a}\\
S_{fyf'y}&=&\frac{fu_0}{\gamma-2i\delta}|e_\varphi|^2.\label{x36b}
\end{eqnarray}
\end{subequations}
Here,  we have introduced the notations $e_r$, $e_{\varphi}$, and $e_z$ for the cylindrical components of the guided-mode profile function $\mathbf{e}(r,\varphi,z)$
of the forward counterclockwise quasicircularly polarized guided modes. The explicit expressions for $e_r$, $e_{\varphi}$, and $e_z$ 
are given in Refs.~\cite{fiber books,fibermode,cesium decay} and are summarized in Appendix \ref{sec:guided}. 
We have also introduced the notation 
\begin{equation}\label{x37} 
u_0=\frac{2\omega_LD_{FF'}^2}{3(2F+1)\epsilon_0\hbar v_g}, 
\end{equation}
with $D_{FF'}$ being the reduced matrix element of the electric dipole operator 
for the atomic transitions between the hyperfine levels $F$ and $F'$ of the ground and excited states, respectively [see Eq.~\eqref{v14}].
Equation \eqref{x36a} shows that the scattering coefficients $S_{fxf'x}$ with  $f'=f$ and $f'=-f$ are proportional to $|e_r|^2+|e_z|^2$ and $|e_r|^2-|e_z|^2$, respectively.
This leads to the difference between the forward and backward scattering in the case where the guided probe field is quasilinearly polarized 
along the major principal axis $x$. Meanwhile, Eq.~\eqref{x36b} shows that the scattering coefficients  $S_{fyf'y}$ do not depend on the propagation direction $f'$ of the guided probe field. Thus, the scattering amplitudes for the forward and backward directions have the same magnitude when the guided probe field is quasilinearly polarized 
along the minor principal axis $y$. The difference between the magnitudes of $S_{fxf'x}$ and $S_{fyf'y}$ leads to the possibility of a manifestation of birefringence \cite{Dawkins11}.

\subsection{Field-transfer matrix and input-output equation}
\label{subsec:input-output}

We first consider an atom with the axial coordinate $z$. We introduce the notations $z^{\pm}=\lim_{\varepsilon\to 0_+} z\pm\varepsilon$ for the limiting points. The fields $\mathcal{A}_{+,p}(z^{-})$ and $\mathcal{A}_{-,p}(z^{+})$
with the propagation directions $f=+$ and $f=-$, respectively, at the left- and right-hand-side limiting points $z^{-}$ and $z^{+}$, respectively, can be interpreted as incoming fields with respect to the atom. 
In contrast, the fields $\mathcal{A}_{+,p}(z^{+})$ and $\mathcal{A}_{-,p}(z^{-})$ with the propagation directions $f=+$ and $f=-$, respectively, at the limiting points $z^{+}$ and $z^{-}$, respectively, can be considered as outgoing fields with respect to the atom. According to the causal principle, the atom interacts with the incoming fields $\mathcal{A}_{\pm,p}(z^{\mp})$
but not with the outgoing fields $\mathcal{A}_{\pm,p}(z^{\pm})$. When we integrate Eq.~\eqref{x33} over an infinitely small interval around the position $z$ of the atom, we obtain 
\begin{eqnarray}\label{x38e}
\mathcal{A}_{fp}(z^{+})&=&\mathcal{A}_{fp}(z^{-})-\sum_{p'}S_{f,p,+,p'}\mathcal{A}_{+,p'}(z^{-})
\nonumber\\&&\mbox{}
-\sum_{p'}S_{f,p,-,p'}\mathcal{A}_{-,p'}(z^{+}).
\end{eqnarray}
We can rewrite the above equation in the matrix form
\begin{equation}\label{xy9}
\boldsymbol{\mathcal{A}}(z^{+})=\mathbf{M}\boldsymbol{\mathcal{A}}(z^{-}),
\end{equation}
where 
\begin{equation}\label{xy10}
\mathbf{M}=(\mathds{1}+\mathbf{S}^{(-)})^{-1}(\mathds{1}-\mathbf{S}^{(+)})
\end{equation}
is the single-atom field-transfer matrix. Here,  we have introduced the matrices $\mathbf{S}^{(+)}$ and $\mathbf{S}^{(-)}$ whose elements are given as 
\begin{eqnarray}\label{xy11}
S^{(+)}_{fpf'p'}&=&S_{fpf'p'}\delta_{f',+},\nonumber\\
S^{(-)}_{fpf'p'}&=&S_{fpf'p'}\delta_{f',-}.
\end{eqnarray}
Note that $\mathbf{S}^{(+)}+\mathbf{S}^{(-)}=\mathbf{S}$. 

We now consider an atomic array.
For convenience, we label the atoms in the array in the order of the increasing axial coordinate $z$. Then, we have $z_1\leq z_2\leq\dots\leq z_N$, where $N$ is the number of atoms in the array. It is clear from Eq.~\eqref{x31} that the scattering matrix $\mathbf{S}=\mathbf{S}(r,\varphi)$ depends on the position $(r,\varphi)$ of the atom in the fiber transverse plane through the coupling coefficients $\mathcal{G}_{fpeg}=\mathcal{G}_{fpeg}(r,\varphi)$. We introduce the notations $\mathbf{M}_j=\mathbf{M}(r_j,\varphi_j)$.
Then, we find that the change in the photon flux amplitude vector $\boldsymbol{\mathcal{A}}$ 
due to atom $j$ is described by the input-output relation
\begin{equation}\label{x38a}
\boldsymbol{\mathcal{A}}(z_j^{+})=\mathbf{M}_j\boldsymbol{\mathcal{A}}(z_j^{-}).
\end{equation}
It follows from Eq.~\eqref{x33} that
the change in the photon flux amplitude vector $\boldsymbol{\mathcal{A}}$ due to the propagation along the fiber in the atom-free path from atom $j$ to atom $j+1$ is given by the formula 
\begin{equation}\label{x38b}
\boldsymbol{\mathcal{A}}(z_{j+1}^{-})=\mathbf{F}_j\boldsymbol{\mathcal{A}}(z_j^{+}),
\end{equation}
where
\begin{equation}\label{x38d}
\mathbf{F}_j=\exp[i\mathbf{B}(z_{j+1}-z_j)]
\end{equation}
is the atom-free guided-field propagator. 
When we combine the relations \eqref{x38a} and \eqref{x38b}, we find the map
\begin{equation}\label{x38}
\boldsymbol{\mathcal{A}}(z_N^{+})=\mathbf{M}_N \mathbf{F}_{N-1}\mathbf{M}_{N-1}\cdots \mathbf{F}_{2}\mathbf{M}_2\mathbf{F}_{1}\mathbf{M}_1\boldsymbol{\mathcal{A}}(z_1^{-}).
\end{equation}

We note that the map \eqref{x38} includes multiple scattering into the guided modes, which propagate along the atomic array in the positive ($+z$) and negative ($-z$) directions of the fiber axis. However, for the effect of scattering into the radiation modes, only single scattering is taken into account in the map. 
Due to the linear geometry of the atomic array, the effect of multiple scattering into the radiation modes is weak and can therefore be neglected.
We note that the transfer matrix formalism has been applied to a one-dimensional optical lattice of two-level atoms in free space \cite{Deutsch95},
a one-dimensional array of atomic layers \cite{Henkel03,Artoni05}, 
an array of two-level atoms along a waveguide with a scalar light field \cite{Chang11,Chang12}, 
an array of coherently driven three-level atoms with a scalar light field \cite{Petrosyan07,Schilke12},
and an array of point-scatterers \cite{Ritsch14a,Ritsch14b}.

We now assume that the atomic array is periodic along the fiber axis, that is, $z_{j+1}-z_j=\Lambda$  for $j=1,\dots,N-1$, while 
$x_j=x_0$ and $y_j=y_0$ for $j=1,\dots,N$. Such a periodic array can be called a grating or a one-dimensional lattice of atoms.
Periodic arrays of atoms along a nanofiber have been experimentally realized in a two-color nanofiber-based atom trap \cite{Vetsch10,Goban12,Mitsch14a}.
For the above-described periodic atomic grating, we have
$\mathbf{F}_j=\mathbf{F}$ and $\mathbf{M}_j=\mathbf{M}$, with 
\begin{eqnarray}\label{x39c}
\mathbf{F}&=& e^{i\mathbf{B}\Lambda}, \nonumber\\
\mathbf{M}&=& \mathbf{M}(r_0,\varphi_0). 
\end{eqnarray}
Equation (\ref{x38}) then takes the input-output form
\begin{equation}\label{x39}
\boldsymbol{\mathcal{A}}_R=\mathbf{W}\boldsymbol{\mathcal{A}}_L,
\end{equation}
where $\boldsymbol{\mathcal{A}}_L=\boldsymbol{\mathcal{A}}(z_1^{-})$ and $\boldsymbol{\mathcal{A}}_R=\boldsymbol{\mathcal{A}}(z_N^{+})$
are the guided-field amplitude vectors at the left- and right-side borders of the atomic array, respectively (see Fig.~\ref{fig1}). 
The notation 
\begin{equation}\label{x39a}
\mathbf{W}=\mathbf{T}^{N-1}\mathbf{M}
\end{equation}
stands for the total (multiatom) field-transfer matrix, with
\begin{equation}\label{x39b}
\mathbf{T}=\mathbf{M}\mathbf{F}
\end{equation}
being the transfer matrix for a single spatial period of the atomic array.

In the basis formed by the guided modes $fp$,
we map the mode indices $(+,p)$, $(+,\bar{p})$, $(-,p)$, and $(-,\bar{p})$ to the numbers $n=1$, 2, 3, and 4, respectively. 
We write $\boldsymbol{\mathcal{A}}_L=(X_1^{(\mathrm{in})},X_2^{(\mathrm{in})},X_3,X_4)$ 
and $\boldsymbol{\mathcal{A}}_R=(X_1,X_2,X_3^{(\mathrm{in})},X_4^{(\mathrm{in})})$. 
Here,  $X_1^{(\mathrm{in})}$ and $X_2^{(\mathrm{in})}$ are the amplitudes of 
the guided fields in the modes $fp$ with the positive propagation direction $f=+$, which are incident onto the atomic array from the left-hand side, and 
$X_3^{(\mathrm{in})}$ and $X_4^{(\mathrm{in})}$ 
are the amplitudes of the guided fields in the modes $fp$ with the negative propagation direction $f=-$, which are incident onto the atomic array from the right-hand side.
Then, we find from Eq.~(\ref{x39}) the following solution:
\begin{eqnarray}\label{x41}
X_3&=&\frac{1}{Q}\Big[W_{34}\Big(W_{41}X_1^{(\mathrm{in})}+W_{42}X_2^{(\mathrm{in})}-X_4^{(\mathrm{in})}\Big)
\nonumber\\&&\mbox{}
-W_{44}\Big(W_{31}X_1^{(\mathrm{in})}+W_{32}X_2^{(\mathrm{in})}-X_3^{(\mathrm{in})}\Big)\Big],
\nonumber\\
X_4&=&\frac{1}{Q}\Big[W_{43}\Big(W_{31}X_1^{(\mathrm{in})}+W_{32}X_2^{(\mathrm{in})}-X_3^{(\mathrm{in})}\Big)
\nonumber\\&&\mbox{}
-W_{33}\Big(W_{41}X_1^{(\mathrm{in})}+W_{42}X_2^{(\mathrm{in})}-X_4^{(\mathrm{in})}\Big)\Big],
\end{eqnarray}
and 
\begin{eqnarray}\label{x42}
X_1&=&W_{11}X_1^{(\mathrm{in})}+W_{12}X_2^{(\mathrm{in})}+W_{13}X_3+W_{14}X_4,
\nonumber\\
X_2&=&W_{21}X_1^{(\mathrm{in})}+W_{22}X_2^{(\mathrm{in})}+W_{23}X_3+W_{24}X_4.
\end{eqnarray}
Here,  we have introduced the notation $Q=W_{33}W_{44}-W_{34}W_{43}$.

We note that, when the lattice constant $\Lambda$ is not close to any integer multiple of the in-fiber half-wavelength $\lambda_F/2=\pi/\beta_L$ of the probe field, 
the atomic array is far off the Bragg resonance. 
In this case, the effect of the interference between the beams reflected from different atoms in the array is not significant and, therefore, 
we can neglect the discreteness and periodicity of the atomic array. 
This approximation means that we can use the atom number distribution $n_A=(1/\Lambda)\delta(x-x_0)\delta(y-y_0)$, which is constant in the axial coordinate $z$.
With the use of this approximation, Eq.~(\ref{x33}) reduces to
\begin{equation}\label{x40}
\frac{\partial }{\partial z}\boldsymbol{\mathcal{A}}=\left(i\mathbf{B} -\frac{\mathbf{S}}{\Lambda}  \right)\boldsymbol{\mathcal{A}}.
\end{equation}
The solution to the above equation is given by Eq.~\eqref{x39} with 
\begin{equation}\label{x40a}
\mathbf{W}=e^{(i\mathbf{B}-\mathbf{S}/\Lambda) L}.
\end{equation}
Here,  we have assumed that $z=0$ and $z=L$ are the left- and right-edge positions of the atomic medium, respectively.

\section{Reflection and transmission of guided light}
\label{sec:quasilinear}

In this section, we study the reflection and transmission of quasilinearly polarized guided light fields.

\subsection{Reflection and transmission coefficients}
\label{subsec:transfer}

We use the basis formed by the guided modes $fx$ and $fy$, which are quasilinearly polarized along the principal directions $x$ and $y$, respectively. In this basis, the cross-polarization scattering matrix elements $S_{fxf'y}$ and $S_{fyf'x}$ are, according to Eq.~\eqref{x35}, zero. This means that the modes with the orthogonal principal polarizations $x$ and $y$ are not coupled to each other by the atoms in the array. Therefore, we can derive a closed set of propagation equations for the guided modes with a single principal quasilinear polarization $\xi=x$ or $\xi=y$. It is convenient to introduce the notations
$\mathcal{A}_{+}=\mathcal{A}_{+,\xi}$ and $\mathcal{A}_{-}=\mathcal{A}_{-,\xi}$ for the field amplitudes of these modes. Then, the amplitudes $\mathcal{A}_{\pm}(N)$ and $\mathcal{A}_{\pm}(0)$ of the fields at the right and left ends of the linear $N$-atom array, respectively, are related to each other by the equation 
\begin{equation}
\left(\begin{array}{c}\mathcal{A}_{+}(N)\\\mathcal{A}_{-}(N)\end{array}\right)
=\mathbf{W}\left(\begin{array}{c}\mathcal{A}_{+}(0)\\\mathcal{A}_{-}(0)\end{array}\right).
\label{x45}
\end{equation}
Here,  the notation $\mathbf{W}=(\mathbf{M}\mathbf{F})^{N-1}\mathbf{M}$
stands for the total transfer matrix, with
\begin{eqnarray}
\mathbf{M}
&=&\left(\begin{array}{cc}1&S_{+-}\\0&1+S_{--}\end{array}\right)^{-1}\left(\begin{array}{cc}1-S_{++}&0\\-S_{-+}&1\end{array}\right)\nonumber\\
&=&\left(\begin{array}{cc}M_{11}&M_{12}\\M_{21}&M_{22}\end{array}\right)
\label{x46}
\end{eqnarray}
being the field transfer matrix of a single atom and
\begin{equation}
\mathbf{F}=\left(\begin{array}{cc}e^{i\beta_L \Lambda}&0\\0&e^{-i\beta_L \Lambda}\end{array}\right)
\label{x47}
\end{equation}
being the propagator for the guided modes of the fiber without atoms. In Eq.~\eqref{x46}, we have used the abbreviation
$S_{ff'}=S_{f\xi f'\xi}$, where $f,f'=+,-$ and $\xi=x,y$, for the scattering matrix elements of the guided modes with a given quasilinear principal polarization $\xi$.
It follows from Eqs.~\eqref{x46} that the matrix elements of the single-atom field transfer matrix $\mathbf{M}$ are 
\begin{eqnarray}\label{x48a}
M_{11}&=&1-S_{++}+\frac{S_{+-}S_{-+}}{1+S_{--}},\nonumber\\
M_{22}&=&\frac{1}{1+S_{--}},\nonumber\\
M_{12}&=&-\frac{S_{+-}}{1+S_{--}},\nonumber\\
M_{21}&=&-\frac{S_{-+}}{1+S_{--}}.
\end{eqnarray}
With the help of Eqs.~\eqref{x36}, we find  that
the matrix elements are given, in the case of the fields with the major principal polarization $\xi=x$, as
\begin{eqnarray}
M_{11}&=&\frac{(1-2S_r)(1-2S_z)}{1-S_r-S_z},\nonumber\\
M_{22}&=&\frac{1}{1-S_r-S_z},\nonumber\\
M_{21}&=&-M_{12}=\frac{S_r-S_z}{1-S_r-S_z},
\label{x48}
\end{eqnarray}
and, in the case of the fields with the minor principal polarization $\xi=y$, as
\begin{eqnarray}
M_{11}&=&\frac{1-2S_\varphi}{1-S_\varphi},\nonumber\\
M_{22}&=&\frac{1}{1-S_\varphi},\nonumber\\
M_{21}&=&-M_{12}=\frac{S_\varphi}{1-S_\varphi}.
\label{x49}
\end{eqnarray}
Here,  we have introduced the notations
\begin{subequations}\label{x50}
\begin{eqnarray}
S_r&=&\frac{u_0}{\gamma-2i\delta}|e_r|^2,\label{x50a}\\
S_\varphi&=&\frac{u_0}{\gamma-2i\delta}|e_\varphi|^2,\label{x50b}\\
S_z&=&\frac{u_0}{\gamma-2i\delta}|e_z|^2.\label{x50c}
\end{eqnarray}
\end{subequations}
Note that $M_{11}M_{22}-M_{12}M_{21}=1$.

The reflection and transmission coefficients of a single atom are given by
$R=-M_{21}/M_{22}$ and $T=1/M_{22}$, respectively.
The explicit expressions for these coefficients are found, for the field with the polarization $\xi=x$, to be
\begin{eqnarray}\label{x51}
R&=&-S_r+S_z,\nonumber\\
T&=&1-S_r-S_z,
\end{eqnarray}
and, for the field with the polarization $\xi=y$, to be
\begin{subequations}\label{x52}
\begin{eqnarray}
R&=&-S_\varphi,\label{x52a}\\
T&=&1-S_\varphi.\label{x52b}
\end{eqnarray}
\end{subequations}
It is clear that the coefficients $R$ and $T$ depend on the principal polarization $\xi=x,y$ of the guided light field. 

Note that, according to Eqs.~\eqref{x52}, the relation $T=1+R$ is valid in the case of $y$-polarized guided fields, like the cases of plane waves \cite{Deutsch95} and scalar guided fields \cite{Chang11,Chang12}. However, according to Eqs.~\eqref{x51}, we have $T\not=1+R$ in the case of $x$-polarized guided fields. The difference between the transmission coefficient $T$ and 
the coefficient $1+R$ is associated with the asymmetry between the forward and backward scattering \cite{Fam14}. 
The physical origin of this deviation lies in the complex vector nature of the local polarization of $x$-polarized guided fields, which have a longitudinal component $e_z$ with a relative phase of $\pi/2$ with respect to the transverse radial component $e_r$ [see Eqs.~\eqref{g2} and \eqref{g9}].

Furthermore, comparison between Eqs.~\eqref{x31} and \eqref{x36b} yields $u_0|e_\varphi|^2=\gamma^{(y)}_{\mathrm{1D}}$. Here,  $\gamma^{(y)}_{\mathrm{1D}}=\sum_{feg} \gamma_{eg}^{(fy)}/(2F+1)$ is a measure of the rate of decay into the $y$-polarized guided modes, where $\gamma_{eg}^{(fy)}$ is given by Eq.~\eqref{v3b}. Then, Eq.~\eqref{x50b} becomes 
$S_\varphi=\gamma^{(y)}_{\mathrm{1D}}/(\gamma-2i\delta)$. Hence, Eq.~\eqref{x52a} indicates that,  in the case of $y$-polarized guided fields, the single-atom reflection coefficient is 
$R=-\gamma^{(y)}_{\mathrm{1D}}/(\gamma-2i\delta)$, in agreement with Ref.~\cite{Deutsch95}. 

By diagonalizing the single-period transfer matrix $\mathbf{T}=\mathbf{M}\mathbf{F}$, we find that
the explicit expressions for the elements of the total transfer matrix $\mathbf{W}$ are 
\begin{eqnarray}
W_{11}&=&M_{11}\frac{\sinh (N\theta)}{\sinh\theta}
-e^{-i\beta_L \Lambda}\frac{\sinh [(N-1)\theta]}{\sinh\theta},
\nonumber\\
W_{22}&=&M_{22}\frac{\sinh (N\theta)}{\sinh\theta}
-e^{i\beta_L \Lambda}\frac{\sinh [(N-1)\theta]}{\sinh\theta},
\nonumber\\
W_{21}&=&-W_{12}=M_{21}\frac{\sinh (N\theta)}{\sinh\theta},
\label{x53}
\end{eqnarray}
where
\begin{equation}\label{x54} 
\theta=\ln (D\pm \sqrt{D^2-1}),
\end{equation} 
with
$D=\frac{1}{2}\big(M_{11}e^{i\beta_L \Lambda}+M_{22}e^{-i\beta_L \Lambda}\big)$.
Note that $\theta$ is, in general, a complex number and satisfies the relations $\cosh\theta=D$ and $\sinh\theta=\pm\sqrt{D^2-1}$.

The reflection and transmission coefficients of the atomic array are given by
$R_N=-W_{21}/W_{22}$ and $T_N=1/W_{22}$, respectively. The explicit expressions for these coefficients are found to be
\begin{eqnarray}\label{x55}
R_N&=&\frac{R\sinh (N\theta)}{\sinh (N\theta)-Te^{i\beta_L \Lambda}\sinh [(N-1)\theta]},\nonumber\\
T_N&=&\frac{T\sinh\theta}{\sinh (N\theta)-Te^{i\beta_L \Lambda}\sinh [(N-1)\theta]}.
\end{eqnarray}
We note that  $R_N$ and $T_N$ satisfy the recurrence formulas
\begin{eqnarray}
R_{N+1}&=&R_N+\frac{T_N^2Re^{2i\beta_L \Lambda}}{1-R_NRe^{2i\beta_L \Lambda}},
\nonumber\\
T_{N+1}&=&\frac{T_N Te^{i\beta_L \Lambda}}{1-R_NRe^{2i\beta_L \Lambda}},
\label{x56}
\end{eqnarray}
which are in agreement with ray optics. Like the single-atom reflection and transmission coefficients $R$ and $T$, the corresponding multi-atom coefficients $R_N$ and $T_N$ depend on the principal polarization $\xi=x,y$ of the guided light field. 
In addition, since the parameter $\theta$ is, in general, a complex number, the coefficients $R_N$ and $T_N$ may oscillate when the atom number $N$ or the array period $\Lambda$ varies.

The reflectivity and transmittivity of the atomic grating are given by $|R_N|^2$ and $|T_N|^2$, respectively. A Bragg resonance occurs when the period of the grating and the propagation constant of the light field are such that the reflectivity $|R_N|^2$ achieves a local maximum. When the field phase shift caused by a single atom is small, the Bragg resonance is approximately determined by the geometric Bragg condition $\beta_L \Lambda=n\pi$, where $n=1,2,\dots$ is a positive integer number. Under this condition, the waves reflected from different atoms in the array are almost in phase and, therefore, the reflectivity $|R_N|^2$ is expected to achieve a local maximum.  

In the case of lossless Bragg gratings, we have the equality $|R_N|^2+|T_N|^2=1$. In this case, the transmittivity $|T_N|^2$ has a local minimum at the Bragg resonance, where the reflectivity $|R_N|^2$ has a local maximum. However, in the case of periodic atomic arrays considered here, due to the presence of spontaneous emission into the radiation modes, we have the inequality $|R_N|^2+|T_N|^2<1$. Due to this fact, both the transmittivity $|T_N|^2$ and the reflectivity $|R_N|^2$ may have local peaks at the Bragg resonance (see Fig.~\ref{fig11} and the discussion around this figure in the next section).

\subsection{Bragg resonance}
\label{subsec:Bragg}

Let us analyze the specific case where the geometric Bragg resonance condition is satisfied, that is, $\beta_L\Lambda=n\pi$, with $n=1,2,\dots$ being the order of the Bragg resonance. In this case, with the use of the transformation $\theta=\vartheta+in\pi$, we can rewrite Eqs.~\eqref{x55} as
\begin{subequations}\label{x57}
\begin{eqnarray}
R_N&=&\frac{R\sinh (N\vartheta)}{\sinh (N\vartheta)-T\sinh [(N-1)\vartheta]},\label{x57a}\\
T_N&=&(-1)^{(N+1)n}\frac{T\sinh\vartheta}{\sinh (N\vartheta)-T\sinh [(N-1)\vartheta]}.\qquad \label{x57b}
\end{eqnarray}
\end{subequations}
In the case of $x$ polarization, the parameter $\vartheta$ is found to be  
$\vartheta=\vartheta_r+i\vartheta_i\simeq 2\gamma_s/(\gamma-2i\delta)$, where
\begin{subequations}\label{x58}
\begin{eqnarray}
\vartheta_r\equiv\mathrm{Re}(\vartheta)&\simeq&\frac{2\gamma_{s}\gamma}{\gamma^2+4\delta^2},\label{x58a}\\
\vartheta_i\equiv\mathrm{Im}(\vartheta)&\simeq&\frac{4\gamma_{s}\delta}{\gamma^2+4\delta^2},\label{x58b}
\end{eqnarray}
\end{subequations}
with 
\begin{equation}\label{x59}
\gamma_{s}=u_0|e_re_z|.
\end{equation}
In the case of $y$ polarization, we have $\vartheta=0$.

We note that the parameter $\gamma_{s}$ originates from the difference between $S_{++}S_{--}$ and $S_{+-}S_{-+}$ in the case of $x$ polarization.
The magnitude of $\gamma_s$ cannot exceed the magnitude of the rate $\gamma^{(\mathrm{gyd})}$
of spontaneous emission into guided modes. Since $\gamma^{(\mathrm{gyd})}$ is essentially smaller than the total decay rate $\gamma$, we have the relation $\gamma_s\ll\gamma$, which leads to $|\vartheta|\ll1$. 

We now consider the $x$- and $y$-polarized guided fields separately.

\subsubsection{Case of $x$-polarized guided fields} 

In the case where the transmitted and reflected guided fields are polarized along the major principal axis $x$, we have 
$\gamma_s=u_0|e_re_z|\not=0$. The fact that $\gamma_s$ is nonzero in this case is a consequence of the difference between the scattering matrix elements 
$S_{fx fx}=S_{ff}$ and $S_{fx \bar{f}x}=S_{f\bar{f}}$, which characterize the forward and backward scattering processes, respectively. The nonvanishing $\gamma_s$ is related to the nonvanishing off-diagonal spontaneous emission coefficients $\gamma_{e,e\pm1}^{(fx)}$,
and is a result of the existence of the longitudinal component $e_z$ of the mode profile function 
$\mathbf{e}^{(\omega fx)}$. Since $\gamma_s\not=0$, the real part $\vartheta_r\equiv\mathrm{Re}(\vartheta)$ [see Eq.~\eqref{x58a}] is nonzero and the imaginary part $\vartheta_i\equiv\mathrm{Im}(\vartheta)$ [see Eq.~\eqref{x58b}] depends on the field detuning $\delta$. When $\delta\not=0$, we have $\vartheta_i\not=0$. This leads to modulations of the function $\sinh (N\vartheta)$ when $N$ varies. Furthermore, due to the dependence of $\vartheta_i$ on $\delta$, the function $\sinh (N\vartheta)$ oscillates when $\delta$ varies. 
Consequently, the reflectivity $|R_N|^2$, which is governed by the factor $|\sinh (N\vartheta)|^2$ [see Eq.~\eqref{x57a}], may modulate when $N$ or $\delta$ varies. The nonzero value of $\vartheta_r$ leads to a decreasing behavior of the envelope of the modulations of $|R_N|^2$ when $N$ increases at a nonzero detuning $\delta$ [see the dashed red curve of Fig.~\ref{fig8}(b) and the discussion around this figure in the next section]. A simple qualitative estimate shows that the strength of the reduction of the envelope is proportional to the factor 
$N\vartheta_r\simeq 2\gamma_{s}\gamma N/(\gamma^2+4\delta^2)$. When $N\vartheta_r\gg1$, that is, when $2\gamma_{s}\gamma N\gg \gamma^2+4\delta^2$, 
we can use the approximation $\sinh (N\vartheta)\simeq e^{N\vartheta}/2$. When we apply this approximation to Eqs.~\eqref{x57}, we obtain
\begin{equation}\label{x60}
\begin{split}
R_N&\to R_\infty=\frac{R}{1-Te^{-\vartheta}},\\
T_N&\to 0.
\end{split}
\end{equation}
With the help of Eqs.~\eqref{x51} and \eqref{x58}, we find
\begin{equation}\label{x61}
R_\infty\simeq -\frac{|e_r|-|e_z|}{|e_r|+|e_z|}.
\end{equation}
It is clear that the limiting value $R_\infty$ is determined by the guided-mode profile functions $e_r$ and $e_z$ only.
It does not depend on the other parameters. In particular, $R_\infty$ does not depend on the detuning $\delta$.
Thus, when $N$ is large enough and $\delta$ is small enough, under the Bragg resonance condition, the modulations of $|R_N|^2$ are suppressed and the magnitude of $|R_N|^2$ does not vary significantly and approaches a constant value given by Eq.~\eqref{x61} [see the asymptotic behavior of the curves in Fig.~\ref{fig8}(b), the plateaus in Figs.~\ref{fig12}(b) and \ref{fig13}(a), the central plateau in Fig.~\ref{fig14}(a), and the discussions around these figures in the next section].
Since $|e_r|> |e_z|>0$, we have $|R_\infty|^2<1$, that is, the limiting value $|R_\infty|^2$ of the reflectivity for the $x$-polarized guided fields is strictly smaller than unity. Since the limiting value $|R_\infty|^2$ does not depend on the detuning $\delta$, a plateau may appear in the frequency dependence of the reflectivity $|R_N|^2$
when $N$ is large enough [see Fig.~\ref{fig13}(a)]. The condition for the appearance of such a plateau is $N\gg\gamma/\gamma_s$.
The characteristic values of the edges of this plateau are $\pm\delta_{\mathrm{flat}}$. When $N\ll v_g/\gamma\Lambda$,
we obtain the estimate $\delta_{\mathrm{flat}}\simeq\sqrt{\gamma_{s}\gamma N}/2$.

\subsubsection{Case of $y$-polarized guided fields} 

In the case where the transmitted and reflected guided fields are polarized along the minor principal axis $y$, 
we have $S_{fy fy}=S_{fy \bar{f}y}$, that is, $S_{f f}=S_{f \bar{f}}$.  This leads to $\vartheta=0$. 
Then, the application of l'Hospital's rule to Eqs.~\eqref{x57} yields
\begin{eqnarray}\label{x65}
R_N&=&\frac{NR}{1-(N-1)R},\nonumber\\
T_N&=&(-1)^{(N+1)n}\frac{T}{1-(N-1)R},
\end{eqnarray}
With the use of Eqs.~\eqref{x52} for $R$ and $T$, we find
\begin{eqnarray}\label{x66}
R_N&=&-\frac{Nu_0|e_\varphi|^2}{\gamma-2i\delta+(N-1)u_0|e_\varphi|^2},\nonumber\\
T_N&=&(-1)^{(N+1)n}\frac{\gamma-2i\delta-u_0|e_\varphi|^2}{\gamma-2i\delta+(N-1)u_0|e_\varphi|^2}.
\end{eqnarray}
It is clear that, when $N$ or $\delta$ varies, no modulations of $|R_N|^2$ are observed in the case of
the  $y$ polarization, unlike in the case of $x$ polarization. According to the first expression in Eqs.~\eqref{x66},
the reflectivity $|R_N|^2$ under the Bragg resonance condition $\beta_L\Lambda=n\pi$ is a Lorentzian function of the atom-field detuning $\delta$. 
The peak value of this function is $\{Nu_0|e_\varphi|^2/[\gamma+(N-1)u_0|e_\varphi|^2]\}^2$ and
is achieved at the exact atomic resonance $\delta=0$. 
The linewidth is $\gamma+(N-1)u_0|e_\varphi|^2$ and increases with increasing $N$.

In the limit $N\to\infty$, we have 
\begin{equation}\label{x66a}
R_N\to-1
\end{equation}
and $T_N\to0$, that is,
$|R_N|^2\to1$ and $|T_N|^2\to0$ (under the condition $R\not=0$).
This result means that the atomic array under the Bragg resonance condition can act as a perfect mirror for the $y$-polarized guided light fields
in the limit of an infinitely large number of atoms ($N\to\infty$)  [see Figs.~\ref{fig15}(a) and \ref{fig16}(b) and the discussions around these figures]. The loss due to the scattering into the radiation modes is suppressed due to the collective enhancement of scattering into the backward guided modes.

In the limit $NR\ll1$, Eqs.~\eqref{x65} yield $R_N\simeq NR+N(N-1)R^2$ and $T_N\simeq (-1)^{(N+1)n}[T+(N-1)RT+(N-1)^2R^2T]$.
The last terms in these expressions contain $N^2$.  They are signatures of the collective effects.

We note that the Bragg resonance condition $\beta_L\Lambda=n\pi$, considered in this subsection, involves the frequency $\omega_L$ of the guided light field. This condition is slightly different from the condition $\beta_0\Lambda=n\pi$  \cite{Chang12}. 
When we vary the frequency $\omega_L$ of the guided light field in a finite range but fix all the other parameters, the condition $\beta_L\Lambda=n\pi$ is broken and, consequently,
the dependence of the reflectivity $|R_N|^2$ on the field detuning is not strictly a Lorentzian function.
At the exact atomic resonance ($\omega_L=\omega_0$), the conditions $\beta_L\Lambda=n\pi$ and $\beta_0\Lambda=n\pi$ are equivalent to each other.

\subsection{Band gaps}
\label{subsec:gaps}

We consider the case where the frequency $\omega_L$ of the guided probe field may be detuned from but is still near to a Bragg resonance.
In this case, we can write $\beta_L\Lambda=n\pi+\phi$, where $n=1,2,\dots$ is the order of the Bragg resonance and $|\phi|\ll1$
is a small quantity characterizing the mismatch between the spatial period of the guided probe field and that of the atomic array. 
We introduce the lattice Bragg resonance frequency $\omega_{\mathrm{lat}}$ that 
is determined by the equation $\beta(\omega_{\mathrm{lat}})\Lambda=n\pi$.
Then, we have $\phi\simeq (\omega_L-\omega_{\mathrm{lat}})\Lambda/v_g$. 
The complex parameter $\theta$, defined by Eq.~\eqref{x54}, can be presented in the form $\theta=\vartheta+in\pi$,
where the parameter $\vartheta$ is determined by the equation
\begin{equation}\label{x71}
\cosh\vartheta=\frac{M_{11}+M_{22}}{2}\cos\phi+\frac{M_{11}-M_{22}}{2}i\sin\phi.
\end{equation}
Since $|M_{11}-1|\ll1$, $|M_{22}-1|\ll1$, and $|\phi|\ll1$, we have $|\vartheta|\ll1$.
When we expand Eq.~\eqref{x71} into a Taylor series and keep only terms up to second order, we find 
\begin{equation}\label{x72}
\vartheta\simeq\sqrt{M_{11}+M_{22}-2+i(M_{11}-M_{22})\phi-\phi^2}.
\end{equation}
In deriving the above expression we have chosen, without loss of generality, the solution $\vartheta$ with $\mathrm{Re}(\vartheta)>0$
or with $\mathrm{Re}(\vartheta)=0$ and $\mathrm{Im}(\vartheta)\geq0$.

The parameter $-i\vartheta$ characterizes the quasimomentum of the Bloch states of 
the infinite periodic atomic lattice \cite{Deutsch95}.
The band-gap frequencies are defined as the frequencies that lead to, even in the absence of scattering losses, a nonzero imaginary part of the quasimomentum, that is, to a nonzero real part of $\vartheta$.
In the band gap region, the reflectivity and transmittivity coefficients of the atomic array in the limit $N\to\infty$ are found from Eqs.~\eqref{x55} to be
\begin{equation}\label{x73}
R_\infty=\frac{R}{1-Te^{i\phi-\vartheta}}
\end{equation}
and $T_\infty=0$.
Since $|\phi|\ll1$, $|\vartheta|\ll1$, and $T\simeq1$, expression \eqref{x73} can be approximated by 
$R_\infty\simeq R/(1-T-i\phi+\vartheta)$.

Below we consider the cases of $x$- and $y$-polarized guided fields separately.

\subsubsection{Case of $x$-polarized guided fields} 

In the case of $x$-polarized guided probe fields, the matrix elements $M_{11}$ and $M_{22}$ are given by 
Eqs.~\eqref{x48}. In this case, Eq.~\eqref{x72} gives 
\begin{equation}\label{x74}
\vartheta\simeq \sqrt{4S_rS_z-2i(S_r+S_z)\phi-\phi^2}.
\end{equation}
When we neglect $\mathrm{Re}(S_r)$ and $\mathrm{Re}(S_z)$, which are associated with the scattering loss [see Eqs.~\eqref{x50}], 
we obtain the condition $-4\mathrm{Im}(S_r)\mathrm{Im}(S_z)+2\mathrm{Im}(S_r+S_z)\phi-\phi^2>0$ for the band-gap appearance.
This condition requires either $2\mathrm{Im}(S_z)<\phi<2\mathrm{Im}(S_r)$ or $2\mathrm{Im}(S_r)<\phi<2\mathrm{Im}(S_z)$.
Note that $\phi\simeq (\delta-\delta_{\mathrm{lat}})\Lambda/v_g$, where $\delta_{\mathrm{lat}}=\omega_{\mathrm{lat}}-\omega_0$. 
Hence, with the help of expressions~\eqref{x50a} and \eqref{x50c} for $S_r$ and $S_z$, respectively, we find the band gap conditions
\begin{eqnarray}\label{x75}
&&u_0|e_z|^2\frac{4\delta}{\gamma^2+4\delta^2}<\frac{\Lambda}{v_g}(\delta-\delta_{\mathrm{lat}})<u_0|e_r|^2\frac{4\delta}{\gamma^2+4\delta^2},\nonumber\\
&&u_0|e_r|^2\frac{4\delta}{\gamma^2+4\delta^2}<\frac{\Lambda}{v_g}(\delta-\delta_{\mathrm{lat}})<u_0|e_z|^2\frac{4\delta}{\gamma^2+4\delta^2}.\qquad
\end{eqnarray}
Note that $u_0|e_r|^2$ and $u_0|e_z|^2$ are on the order of the characteristic rate $\gamma^{(\mathrm{gyd})}$ of spontaneous emission from an atom into the guided modes,
while $v_g/\Lambda$ is on the order of the atomic frequency $\omega_0$.
When the atoms are not too far away from the fiber and $\delta_{\mathrm{lat}}$ is not too large, we have 
\begin{equation}\label{x77}
\sqrt{u_0|e_{r,z}|^2v_g/\Lambda}\gg\gamma,|\delta_{\mathrm{lat}}|.
\end{equation}
Under this condition, the edges of the band gaps are far from the atomic resonance. Hence, we can neglect $\gamma$ when evaluating the
edges. Then, we find two band gaps that are positioned symmetrically with respect to the middle point $\omega_c=(\omega_0+\omega_{\mathrm{lat}})/2$ between the atomic frequency $\omega_0$ and the lattice frequency $\omega_{\mathrm{lat}}$. 
The two band gaps extend over the frequency range from $\omega_c-\Delta_{\mathrm{max}}$ to $\omega_c-\Delta_{\mathrm{min}}$ and
from $\omega_c+\Delta_{\mathrm{min}}$ to $\omega_c+\Delta_{\mathrm{max}}$. 
Here,  we have introduced the notations
\begin{eqnarray}\label{x78}
\Delta_{\mathrm{max}}&=&\sqrt{\frac{\delta_{\mathrm{lat}}^2}{4}+\frac{u_0|e_r|^2v_g}{\Lambda}},
\nonumber\\
\Delta_{\mathrm{min}}&=&\sqrt{\frac{\delta_{\mathrm{lat}}^2}{4}+\frac{u_0|e_z|^2v_g}{\Lambda}}.
\end{eqnarray}
Note that, since $|e_r|>|e_z|$, we have $\Delta_{\mathrm{max}}>\Delta_{\mathrm{min}}$.
The two band gaps have the same width
\begin{equation}\label{x79}
\Delta_{\mathrm{gap}}=\sqrt{\frac{\delta_{\mathrm{lat}}^2}{4}+\frac{u_0|e_r|^2v_g}{\Lambda}}-\sqrt{\frac{\delta_{\mathrm{lat}}^2}{4}+\frac{u_0|e_z|^2v_g}{\Lambda}}.
\end{equation}

The band gaps will be formed when the number of atoms $N$ is sufficiently large that $N\mathrm{Re}(\vartheta_{\mathrm{gap}})\gg1$, where $\vartheta_{\mathrm{gap}}$ is the characteristic value of $\vartheta$ in the band gap region. Hence,
the threshold value of the atom number required to create band gaps is $N_{\mathrm{gap}}=1/\mathrm{Re}(\vartheta_{\mathrm{gap}})$. Under the condition \eqref{x77}, 
we find that, in the band gap region, $\vartheta$ is approximately a real parameter, with the peak value 
$\vartheta\simeq \sqrt{u_0\Lambda/v_g}(|e_r|-|e_z|)\equiv \vartheta_{\mathrm{gap}}$, achieved at
the detunings $\delta\simeq\pm\sqrt{u_0|e_re_z|v_g/\Lambda}\equiv \pm \delta_{\mathrm{mid}}$. 
Using this estimate for $\vartheta_{\mathrm{gap}}$, we obtain $N_{\mathrm{gap}}=\sqrt{v_g/u_0\Lambda}/(|e_r|-|e_z|)$.
Furthermore, the characteristic reflectivity coefficient in the band gap region is found to be
\begin{eqnarray}\label{x80}
R_{\mathrm{gap}}&\equiv& R_\infty(\pm\delta_{\mathrm{mid}})\simeq -\frac{\sqrt{|e_r|}\pm i\sqrt{|e_z|}}{\sqrt{|e_r|}\mp i\sqrt{|e_z|}}\nonumber\\&&\times
\left(1-\frac{\gamma}{2\sqrt{u_0v_g/\Lambda}(|e_r|-|e_z|) }\right).
\end{eqnarray}
Our numerical calculations for the fiber with the radius $a=250$ nm and for a guided light field with the wavelength $\lambda=852$ nm show that, outside the fiber, we have $2.1>|e_r|/|e_z|>1.75$. This leads to $|e_r|-|e_z|\sim |e_r|\sim|e_z|$.
Hence, under the condition \eqref{x77}, we can neglect the second term in the second line of Eq.~\eqref{x80} and obtain $|R_{\mathrm{gap}}|^2\simeq 1$.

\subsubsection{Case of $y$-polarized guided fields} 

In the case of $y$-polarized guided probe fields, the matrix elements $M_{11}$ and $M_{22}$ are given by 
Eqs.~\eqref{x49}. In this case,  Eq.~\eqref{x72} yields
\begin{equation}\label{x81}
\vartheta\simeq \sqrt{-2iS_\varphi\phi-\phi^2}.
\end{equation}
When we neglect $\mathrm{Re}(S_\varphi)$, which corresponds to the scattering loss,
we obtain the condition $\phi[2\mathrm{Im}(S_\varphi)-\phi]>0$ for the band-gap appearance.
This condition requires either $2\mathrm{Im}(S_\varphi)<\phi<0$ or $0<\phi<2\mathrm{Im}(S_\varphi)$.
Using the expression $\phi\simeq (\delta-\delta_{\mathrm{lat}})\Lambda/v_g$ and expression~\eqref{x50b} for $S_\varphi$, 
we find the band gap conditions
\begin{eqnarray}\label{x82}
&&u_0|e_\varphi|^2\frac{4\delta}{\gamma^2+4\delta^2} < \frac{\Lambda}{v_g}(\delta-\delta_{\mathrm{lat}}) < 0, \nonumber\\
&&0<\frac{\Lambda}{v_g}(\delta-\delta_{\mathrm{lat}}) < u_0|e_\varphi|^2\frac{4\delta}{\gamma^2+4\delta^2}.
\end{eqnarray}
When the atoms are not too far away from the fiber and $\delta_{\mathrm{lat}}$ is not too large, we have 
\begin{equation}\label{x84}
\sqrt{u_0|e_{\varphi}|^2v_g/\Lambda}\gg\gamma,|\delta_{\mathrm{lat}}|.
\end{equation}
Under this condition, the outer edges of the band gaps are far from the atomic resonance. 
Hence, we can neglect $\gamma$ in evaluating the outer edges. Meanwhile, the lattice frequency $\omega_{\mathrm{lat}}$
is always an inner edge, and, for $\omega_{\mathrm{lat}}\not=\omega_0$, there is another inner edge that almost coincides with the atomic frequency $\omega_0$. 
Then, we obtain two band gaps that 
are positioned symmetrically with respect to the middle point $\omega_c$ between the atomic frequency $\omega_0$ and the lattice frequency $\omega_{\mathrm{lat}}$. 
They extend over the frequency range from $\omega_c-\Delta_{\mathrm{max}}$ to $\omega_c+\Delta_{\mathrm{max}}$, with frequencies between the atomic frequency $\omega_0$ and  
the lattice frequency $\omega_{\mathrm{lat}}$ excluded \cite{Deutsch95}. 
Here,  we have introduced the notation
\begin{equation}\label{x85}
\Delta_{\mathrm{max}}=\sqrt{\frac{\delta_{\mathrm{lat}}^2}{4}+\frac{u_0|e_\varphi|^2v_g}{\Lambda}}.
\end{equation}
The two band gaps have the same width
\begin{equation}\label{x86}
\Delta_{\mathrm{gap}}=\sqrt{\frac{\delta_{\mathrm{lat}}^2}{4}+\frac{u_0|e_\varphi|^2v_g}{\Lambda}}
-\frac{|\delta_{\mathrm{lat}}|}{2}.
\end{equation}

The band gaps will be formed when $N\gg N_{\mathrm{gap}}\equiv 1/\mathrm{Re}(\vartheta_{\mathrm{gap}})$, where $\vartheta_{\mathrm{gap}}$ is the characteristic value of $\vartheta$ in the band gap region.
Under the condition \eqref{x84}, Eq.~\eqref{x81} yields $\vartheta\simeq \sqrt{3u_0|e_{\varphi}|^2\Lambda/4v_g}\equiv \vartheta_{\mathrm{gap}}$
for the middle points $\delta\simeq \pm (1/2) \sqrt{u_0|e_{\varphi}|^2v_g/\Lambda}\equiv \pm \delta_{\mathrm{mid}}$ of the band gaps. Using this estimate for $\vartheta_{\mathrm{gap}}$, we find 
$N_{\mathrm{gap}}=\sqrt{4v_g/3u_0\Lambda}/|e_{\varphi}|$. 
Furthermore, the characteristic reflectivity coefficient in the band gap region is found to be
\begin{eqnarray}\label{x87}
R_{\mathrm{gap}}&\equiv& R_\infty(\pm\delta_{\mathrm{mid}})\simeq 
-\frac{1\pm i\sqrt{3}}{2}\nonumber\\&&\mbox{}\times
\left(1-\frac{\gamma}{\sqrt{3u_0|e_{\varphi}|^2v_g/\Lambda}}\right).
\end{eqnarray}
Due to the condition \eqref{x84}, we can neglect the second term in the second line of Eq.~\eqref{x87} and obtain $|R_{\mathrm{gap}}|^2\simeq 1$.
The above results for $y$-polarized guided fields are in agreement with the results for a one-dimensional optical lattice of two-level atoms in free space \cite{Deutsch95} 
and for an array of two-level atoms along a waveguide with a scalar field \cite{Chang12,Chang14}.

\section{Numerical results}
\label{sec:numerical}

In this section, we present and discuss the results of numerical calculations.
As already stated, we consider the transitions between the hyperfine levels $6S_{1/2}F=4$ and $6P_{3/2}F'=5$ of the $D_2$ line of atomic cesium, with the wavelength $\lambda_0=852$ nm. 
The atom is positioned on the positive side of the axis $x$. 
In the numerical calculations, we use the fiber radius $a=250$ nm. 

We use Eq.~\eqref{x31} to calculate the scattering matrix $\mathbf{S}$, use
Eq.~\eqref{x38} to calculate the photon flux amplitudes $\mathcal{A}_{fp}$, and use the coherent-field approximation 
$\mathcal{N}_{fp}\simeq |\mathcal{A}_{fp}|^2$ to calculate the photon flux $\mathcal{N}_{fp}$.
The propagation power of the guided probe light field is assumed to be  much lower than the saturation power $P_{\mathrm{sat}}=4.4$ pW. Here,  $P_{\mathrm{sat}}$ is estimated as the power of a quasicircularly polarized guided light field that produces the intensity 
$I\equiv c\epsilon_0|\mathcal{E}|^2/2=I_{\mathrm{sat}}$ on the fiber surface, where $I_{\mathrm{sat}}=1.1$ mW/cm$^2$ is
the saturation intensity for a cesium atom with the cyclic transition \cite{coolingbook}.
Our results are valid in the linear, quasistationary, weak-excitation regime, where the interaction time is short enough that the redistribution of populations of the Zeeman sublevels of the ground state is small.

\begin{figure}[tbh]
\begin{center}
  \includegraphics{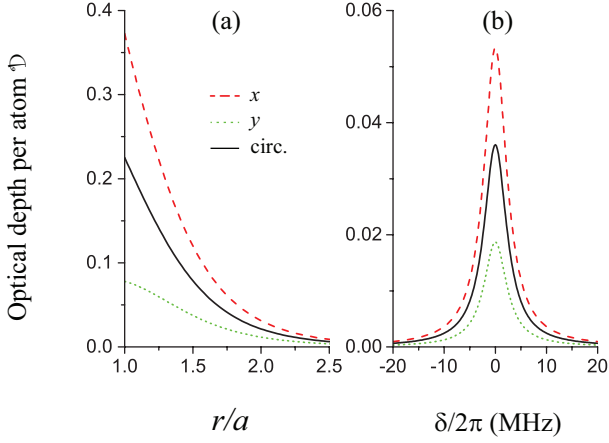}
 \end{center}
\caption{(Color online) 
Dependencies of the optical depth per atom $\mathcal{D}$ on the radial position of the atom (a) and the detuning of the guided probe field (b)  
in the cases where the guided probe field is quasicircularly polarized (solid black lines),
quasilinearly polarized along the $x$ direction (dashed red lines), and quasilinearly polarized along the $y$ direction (dotted green lines). 
The fiber radius is $a=250$ nm and the light wavelength is $\lambda=852$ nm. 
The atom is located on the $x$ axis. 
In (a), the field is tuned to exact resonance with the atom.
In (b), the radial position of the atom is $r/a=1.8$.
}
\label{fig2}
\end{figure}

According to Eq.~\eqref{x32}, the real part of  the diagonal matrix element $S_{fpfp}$ determines the 
optical depth per atom $\mathcal{D}$ for the field of the guided mode $fp$ in the linear, quasistationary, weak-excitation regime.
We plot in Fig.~\ref{fig2} the dependencies of $\mathcal{D}$ on the radial position $r$ of the atom and on the detuning $\delta$ of the guided probe field.
We observe that the optical depth per atom $\mathcal{D}$ can be significant even when the atom is not close to the fiber surface.
For example, for the parameters $r/a=1.8$ and $\delta=0$, we obtain $\mathcal{D}\simeq 0.036$ for the quasicircular polarization
and $\mathcal{D}\simeq 0.053$ and $0.019$ for the $x$ and $y$ polarizations, respectively.
We note that, in the steady-state regime, where the redistribution of populations of the Zeeman sublevels of the ground state has taken place,
the optical depth per atom (equivalent to the efficiency of loss per atom in the context of scattering) can become higher \cite{Fam14}.

%%%%%%%%%%%%%%%%%%%%%%% Figure 2 

\begin{figure}[tbh]
\begin{center}
  \includegraphics{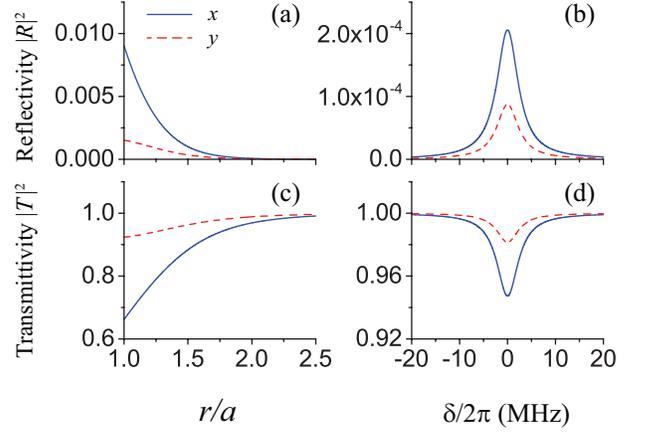}
 \end{center}
\caption{(Color online) 
Dependencies of the single-atom reflectivity $|R|^2$ (upper row) and transmittivity $|T|^2$ (lower row) on the radial position of the atom (left column) and the detuning of the guided probe field (right column)  
in the cases where the guided probe field is 
quasilinearly polarized along the $x$ direction (solid blue lines) and the $y$ direction (dashed red lines). 
The atom is located on the $x$ axis. 
In (a) and (c), the field is tuned to exact resonance with the atom.
In (b) and (d), the radial position of the atom is $r/a=1.8$.
Other parameters are as in Fig.~\ref{fig2}.
}
\label{fig3}
\end{figure}

According to the previous section, the single-atom reflection and transmission coefficients $R$ and $T$, respectively,
are determined by Eqs.~\eqref{x51} and \eqref{x52} for the $x$- and $y$-polarized fields, respectively.
We plot in Fig.~\ref{fig3} the dependencies of the single-atom reflectivity $|R|^2$ and transmittivity $|T|^2$ on the radial position of the atom  and the detuning of the guided probe field. 
We observe that the single-atom reflectivity $|R|^2$ is rather small. The maximum value of $|R|^2$ is about $0.009$, achieved when the field is resonant and $x$ polarized
and the atom is on the fiber surface. 
We note that the magnitude of $|R|^2$ at $\delta=0$ is on the order of $(\gamma^{(\mathrm{gyd})}/\gamma)^2$.
Meanwhile, the single-atom transmittivity $|T|^2$ can be significantly smaller than unity even when the atom is not close to the fiber surface.
This is a result of the significant loss into the radiation modes. We note the approximate relation $1-|T|^2\simeq \mathcal{D}$ between the single-atom transmittivity $|T|^2$
and the optical depth per atom $\mathcal{D}$.

%%%%%%%%%%%%%%%%%%%%%%% Figure 3 

\begin{figure}[tbh]
\begin{center}
  \includegraphics{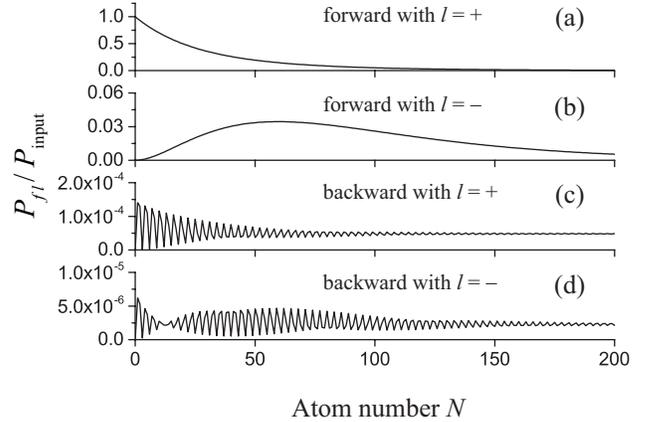}
 \end{center}
\caption{Powers $P_{fl}$ of quasicircularly guided modes $fl$, 
normalized to the input-field power $P_{\mathrm{input}}$, as functions of the number $N$ of atoms in the array.
The input guided probe field propagates along the fiber in the positive direction $+z$ and is counterclockwise ($l=+$) quasicircularly polarized.
The distance from the atoms to the fiber surface is $r-a=200$ nm. The period of the array is $\Lambda=498$ nm. The detuning of the field is $\delta=0$.
Other parameters are as in Fig.~\ref{fig2}. 
}
\label{fig4}
\end{figure}

\begin{figure}[tbh]
\begin{center}
  \includegraphics{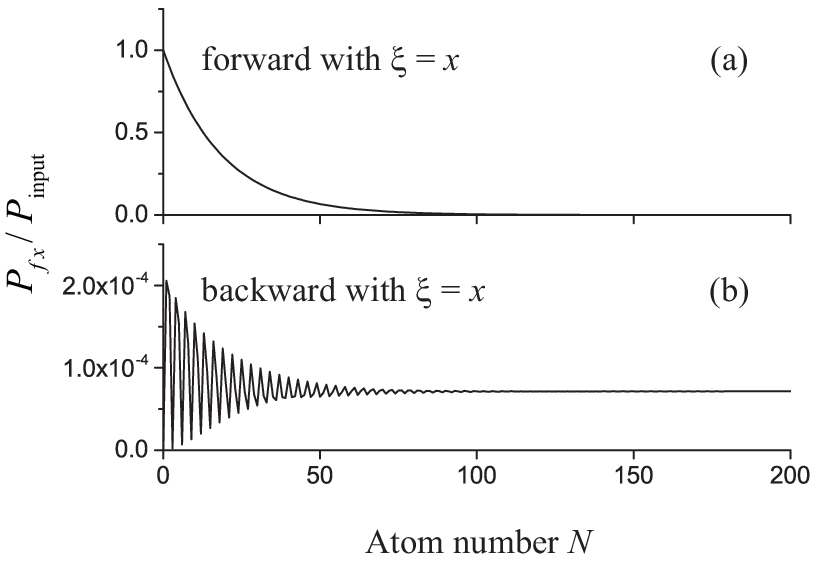}
 \end{center}
\caption{Powers $P_{fx}$ of quasilinearly guided modes $fx$, 
normalized to the input-field power $P_{\mathrm{input}}$, as functions of the number $N$ of atoms in the array.
The input guided probe field propagates along the fiber in the positive direction $+z$ and is quasilinearly polarized along the major principal axis $x$, which is the
radial direction of the atomic position. The distance from the atoms to the fiber surface is $r-a=200$ nm. The period of the array is $\Lambda=498$ nm.  The detuning of the field is $\delta=0$.
Other parameters are as in Fig.~\ref{fig2}. The powers $P_{fy}$ are zero in the considered case and are therefore not plotted.
}
\label{fig5}
\end{figure}

\begin{figure}[tbh]
\begin{center}
  \includegraphics{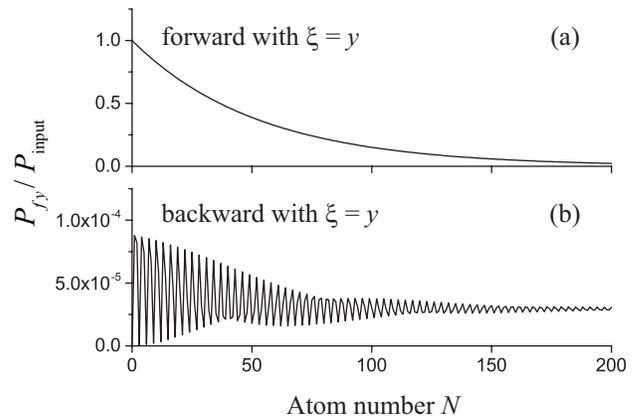}
 \end{center}
\caption{Powers $P_{fy}$ of quasilinearly guided modes $fy$, 
normalized to the input-field power $P_{\mathrm{input}}$, as functions of the number $N$ of atoms in the array.
The input guided probe field propagates along the fiber in the positive direction $+z$ and is quasilinearly polarized along the minor principal axis $y$, which is perpendicular to the
radial direction of the atomic position. The distance from the atoms to the fiber surface is $r-a=200$ nm. The period of the array is $\Lambda=498$ nm.  The detuning of the field is $\delta=0$.
Other parameters are as in Fig.~\ref{fig2}. The powers $P_{fx}$ are zero in the considered case and are therefore not plotted.
}
\label{fig6}
\end{figure}

We plot in Figs.~\ref{fig4}--\ref{fig6} the dependencies of the powers $P_{fp}=\hbar\omega \mathcal{N}_{fp}$
on the number $N$ of atoms in the array in the cases where the probe field is quasicircularly polarized, $x$ polarized, and $y$ polarized, respectively. In the calculations for these figures, we use the array period $\Lambda=498$ nm, which is a half of the in-fiber wavelength
of the red-detuned standing-wave guided light field in the atom trap realized in the experiment \cite{Vetsch10}.  
The powers $P_{fy}$ and $P_{fx}$ are zero in the cases of Figs.~\ref{fig5} and  \ref{fig6}, respectively,
and are therefore not plotted. 
Figures \ref{fig4}(a), \ref{fig5}(a), and \ref{fig6}(a) show that the power of the field in the input mode reduces with increasing atom number $N$. Figure \ref{fig4}(b) shows that, in the case where the input field is counterclockwise (or clockwise) quasicircularly polarized, a new polarization, namely the clockwise (or counterclockwise) quasicircular polarization, is generated during the propagation process. Figures \ref{fig4}(c),  \ref{fig4}(d), \ref{fig5}(b), and \ref{fig6}(b) show that the coherent backward scattering is very weak. In addition, we observe from these figures that the powers of the backward guided modes modulate with increasing number of atoms in the array. Such oscillations are the results of  the interference between the light waves reflected from different atoms in the array. The modulation period is determined by the mismatch between the array period $\Lambda$ and the in-fiber half-wavelength $\lambda_F/2=\pi/\beta_L$ of the probe light.
The irregularities of the oscillations observed in the figures are due to the fact that the number of atoms in the array is a discrete variable.
We note that the normalized powers $P_{f\xi}/P_{\mathrm{input}}$ of the transmitted ($f=+$) and reflected ($f=-$) fields with the quasilinear polarization $\xi=x,y$
are equal to the transmittivity $|T_N|^2$ and the reflectivity $|R_N|^2$, respectively.

%%%%%%%%%%%%%%%%%%%%%%% Figure 4

%%%%%%%%%%%%%%%%%%%%%%% Figure 5

%%%%%%%%%%%%%%%%%%%%%%% Figure 6

\begin{figure}[tbh]
\begin{center}
  \includegraphics{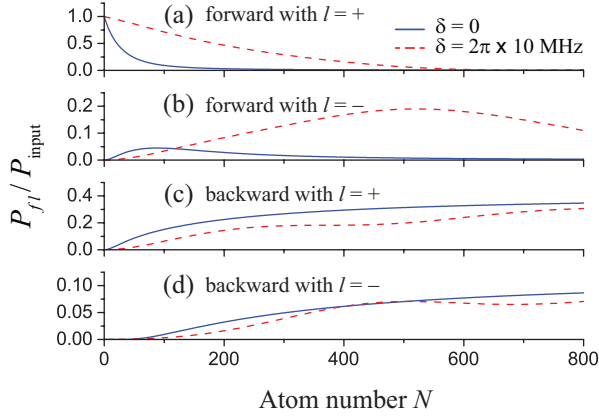}
 \end{center}
\caption{(Color online) Same as Fig.~\ref{fig4} but the period of the array is $\Lambda=2\pi/\beta_L=745$ nm
and the field detuning is $\delta=0$ (solid blue curves) or $2\pi\times 10$ MHz (dashed red curves).
}
\label{fig7}
\end{figure}

\begin{figure}[tbh]
\begin{center}
  \includegraphics{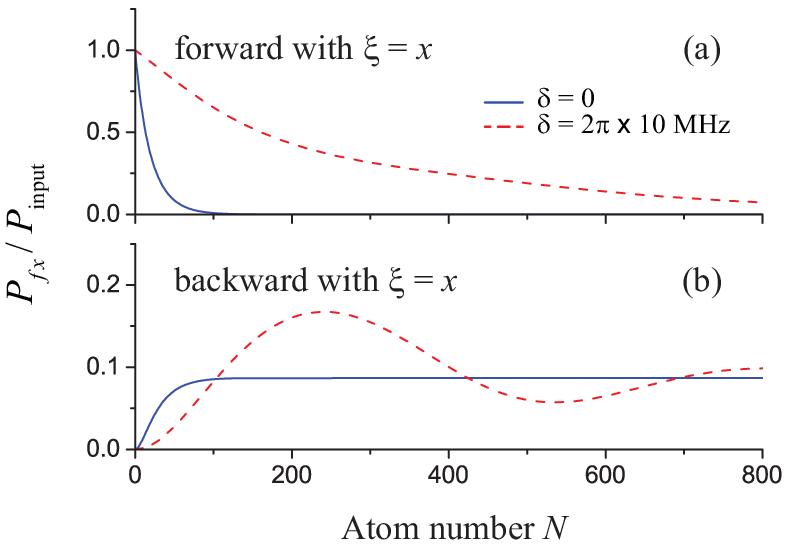}
 \end{center}
\caption{(Color online) Same as Fig.~\ref{fig5} but the period of the array is $\Lambda=2\pi/\beta_L=745$ nm
and the field detuning is $\delta=0$ (solid blue curves) or $2\pi\times 10$ MHz (dashed red curves).
}
\label{fig8}
\end{figure}

\begin{figure}[tbh]
\begin{center}
  \includegraphics{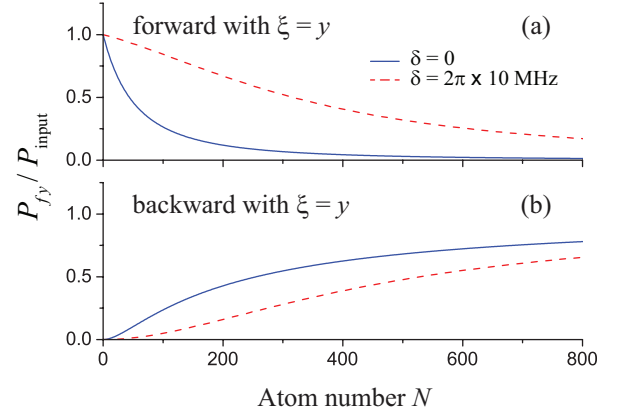}
 \end{center}
\caption{(Color online) Same as Fig.~\ref{fig6}  but the period of the array is $\Lambda=2\pi/\beta_L=745$ nm
and the field detuning is $\delta=0$ (solid blue curves) or $2\pi\times 10$ MHz (dashed red curves).
}
\label{fig9}
\end{figure}

A Bragg resonance occurs when $\beta_L\Lambda$ is an integer multiple of $\pi$. We plot in Figs.~\ref{fig7}--\ref{fig9} the results of calculations for $P_{fp}$ in the case of the Bragg resonance. 
In the framework of our treatment, where the interference between the fields scattered from different atoms into the radiation modes is not taken into account, the magnitudes of the reflectivity $|R_N|^2$
and transmittivity $|T_N|^2$ of the atomic array for the guided fields do not depend on the order of the Bragg resonance. However, to be specific, we take the value $\Lambda=745$ nm, which satisfies the second-order Bragg resonance condition $\beta_L\Lambda=2\pi$  
for the probe field with the atomic resonance frequency $\omega_L=\omega_0$. 
We avoid the first-order Bragg resonance with the aim to minimize the effects of the direct dipole-dipole interaction between the atoms.
We note that the above specific choice of $\Lambda$ corresponds to $\omega_{\mathrm{lat}}=\omega_0$, that is, to $\delta_{\mathrm{lat}}=0$.
We observe from Figs.~\ref{fig7}--\ref{fig9} that the backward scattering becomes significant due to the Bragg resonance. Depending on the polarization and the detuning of the input guided probe field, 
a significant fraction (more than 78\% for $N=800$) of guided light can be scattered into the backward direction [see Fig.~\ref{fig9}(b)]. 
Close inspection of the solid blue lines of Figs.~\ref{fig7}(a), \ref{fig8}(a), and \ref{fig9}(a)
and comparison with Figs.~\ref{fig4}(a), \ref{fig5}(a), and \ref{fig6}(a), respectively, 
reveal that the Bragg resonance condition also leads to an increase in the transmission of the guided probe field.
We note that a significant backward scattering may occur
not only when the guided probe field is at exact resonance with the atoms (solid blue lines) but also when the field is detuned from resonance with the atoms (dashed red lines). Comparison between the dashed red lines (for the case of $\delta/2\pi=10$ MHz) and the solid blue lines (for the case of $\delta=0$) in parts (a) of Figs.~\ref{fig7}--\ref{fig9}
shows that a finite detuning $\delta$ leads to, in general, an increase in the transmission of the guided probe field. 
Such an increase is an obvious result of the reduction of the atomic excitation. 
Comparison between the dashed red line and the solid blue line in  Fig.~\ref{fig8}(b) shows that,
depending on the polarization of the input guided probe field and the number of atoms,
a finite detuning $\delta$ may lead to an increase in the backward scattering. Such an increase
is a result of the competition between the scattering into the backward guided modes on one hand and the scattering into the radiation modes and the forward guided modes on the other hand. 
In the limit $N\to\infty$, the curves in Fig.~\ref{fig8}(b) for the reflected $x$-polarized guided field tend to a constant value $|R_{\infty}|^2\simeq 0.087<1$, in agreement with Eq.~\eqref{x61}.
In the same limit, the curves in Fig.~\ref{fig9}(b) for the reflected $y$-polarized guided field approach unity, in agreement with Eq.~\eqref{x66a}. 
Thus, in the limit $N\to\infty$ and under the condition of the Bragg resonance,
the reflectivity of the $y$-polarized guided field is larger than that of the $x$-polarized guided field.

%%%%%%%%%%%%%%%%%%%%%%% Figure 7

%%%%%%%%%%%%%%%%%%%%%%% Figure 8

%%%%%%%%%%%%%%%%%%%%%%% Figure 9

\begin{figure}[tbh]
\begin{center}
  \includegraphics{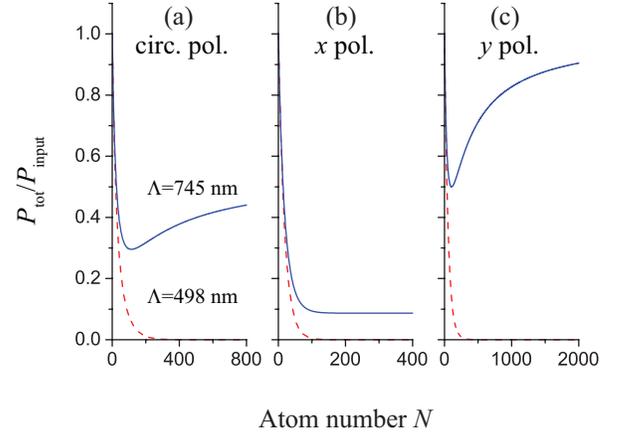}
 \end{center}
\caption{(Color online) Normalized total power $P_{\mathrm{tot}}/P_{\mathrm{input}}$ of the guided fields as a function of the number $N$ of atoms in the array in the cases where the probe field is quasicircularly polarized (a), $x$ polarized (b), and $y$ polarized (c). 
The distance from the atoms to the fiber surface is $r-a=200$ nm. 
The period of the array is $\Lambda=498$ nm (dashed red curves) or $\Lambda=745$ nm (solid blue curves). The field detuning is $\delta=0$. 
Other parameters are as in Fig.~\ref{fig2}. 
}
\label{fig10}
\end{figure}

We plot in Fig.~\ref{fig10} the total power $P_{\mathrm{tot}}=
\sum_{fp}P_{fp}$ as a function of the number $N$ of atoms in the array. 
The dashed red lines, calculated for the case $\Lambda=498$ nm, show that, when the array period is far from the Bragg resonance, 
the total guided-mode power $P_{\mathrm{tot}}$ reduces to a small quantity of the order of $10^{-4}$ with increasing number $N$ of atoms. The solid blue lines, calculated for the case of the Bragg resonance with $\Lambda=2\pi/\beta_L=745$ nm, show that, in this condition, depending on the polarization of the input probe field, $P_{\mathrm{tot}}$ 
may first decrease and then approach a constant nonzero value or rebound with increasing atom number $N$. The rebound of the total guided-mode power indicates the reduction in the scattering into the radiation modes
with increasing number of atoms under the Bragg resonance condition. It is a result of the competition between different channels of scattering. 
In the limit $N\to\infty$, the solid blue curves in Figs.~\ref{fig10}(a), \ref{fig10}(b), and \ref{fig10}(c)
approach the constant nonzero values $0.543$, $0.087$, and 1, respectively. The limiting values $0.087$ and 1 for the solid blue curves in Figs.~\ref{fig10}(b) and \ref{fig10}(c),
respectively, are in agreement with Eqs.~\eqref{x61} and \eqref{x66a}, respectively.
The limiting value $0.543$ for the solid blue curve in Fig.~\ref{fig10}(a) is just the average of the corresponding values for the solid blue curves in Figs.~\ref{fig10}(b) and \ref{fig10}(c).
The reasons are that, on one hand, a quasicircular polarization is a superposition of the $x$ and $y$ polarizations and, on the other hand,
there is no coupling between the $x$- and $y$-polarized guided fields in the linear scattering regime.  
It is clear that the total guided-mode power $P_{\mathrm{tot}}$ in the case of the Bragg resonance (solid blue lines) is larger than that in the nonresonance case (dashed red lines). 
Thus, due to the Bragg resonance condition, the scattering into the radiation modes is reduced.

%%%%%%%%%%%%%%%%%%%%%%% Figure 10

\begin{figure}[tbh]
\begin{center}
  \includegraphics{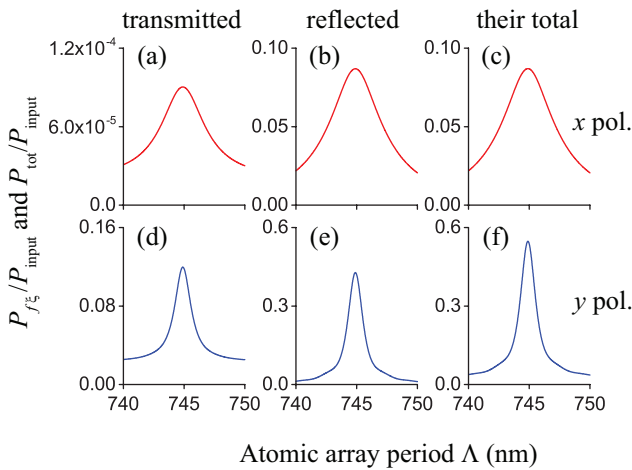}
 \end{center}
\caption{(Color online) Powers $P_{f\xi}$ of the transmitted (left column) and reflected (middle column) guided fields and their total power $P_{\mathrm{tot}}$ (right column),
normalized to the input-field power $P_{\mathrm{input}}$, as functions of 
the array period $\Lambda$. 
The guided probe field is quasilinearly polarized along the major principal direction $x$ (upper row) or 
the minor principal direction $y$ (lower row). The number of atoms in the array is $N=200$. The distance from the atoms to the fiber surface is $r-a=200$ nm. 
The field detuning is $\delta=0$. Other parameters are as in Fig.~\ref{fig2}. 
}
\label{fig11}
\end{figure}

In order to show the effect of the array period $\Lambda$ on the reflection and transmission,
we plot in Fig.~\ref{fig11} the powers $P_{f\xi}$ of the transmitted (left column) and reflected (middle column) quasilinearly polarized guided fields
as well as their total power $P_{\mathrm{tot}}=\sum_f P_{f\xi}$ (right column) as functions of $\Lambda$.
The figure shows that both the transmitted field power and the reflected field power have a local maximum
at the array period $\Lambda=745$ nm, which satisfies the Bragg resonance condition for the field frequency $\omega_L=\omega_0$.
The coexistence of the local maxima of the transmitted field power 
(the trasmittivity $|T_N|^2$) and the reflected field power (the reflectivity $|R_N|^2$) at the Bragg resonance  
is an interesting feature and is a result of the competition between different scattering channels in the presence of loss.
The right column of Fig.~\ref{fig11} shows that, at the Bragg resonance,
the total power $P_{\mathrm{tot}}$ of the transmitted  and reflected guided fields achieves a maximum. 
This result indicates that the scattering from the atoms into the radiation modes is suppressed due to the Bragg resonance,
in agreement with the numerical results presented in Fig.~\ref{fig10}.
It is clear that, in the vicinity of the Bragg resonance, the reflectivity of the $y$-polarized guided field [see Fig.~\ref{fig11}(e)] is larger than that of the $x$-polarized guided field [see Fig.~\ref{fig11}(b)]. The small magnitudes of $P_{f\xi}$ and $P_{\mathrm{tot}}$ in the upper row of Fig.~\ref{fig11} indicate that
most of the energy of the $x$-polarized guided field is scattered into the radiation modes. 
The significant maximal magnitudes of the quantities plotted in Figs.~\ref{fig11}(e) and \ref{fig11}(f) indicate that, in the vicinity of the Bragg resonance, 
most of the energy of the $y$-polarized guided field is reflected.
It is interesting to note that the linewidth of the dependence of the reflectivity on the array period $\Lambda$ is on the order
of a few nanometers. The reason is that, when $\beta_L\Delta\Lambda\ll1$, the Bragg resonance condition $\beta_L \Lambda=n\pi$, with $n=1,2,\dots$, is not significantly deviated.

%%%%%%%%%%%%%%%%%%%%%%% Figure 11

\begin{figure}[tbh]
\begin{center}
  \includegraphics{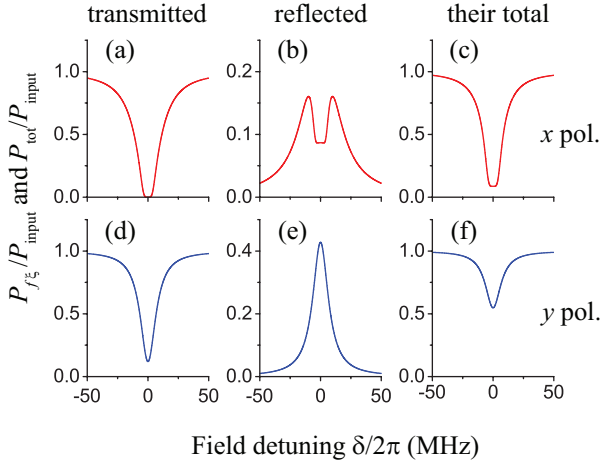}
 \end{center}
\caption{(Color online) Powers $P_{f\xi}$ of the transmitted (left column) and reflected (middle column) guided fields
and their total power $P_{\mathrm{tot}}$ (right column),
normalized to the input-field power $P_{\mathrm{input}}$, as functions of 
the field detuning $\delta$. 
The guided probe field is quasilinearly polarized along the major principal direction $x$ (upper row) or 
the minor principal direction $y$ (lower row). The number of atoms in the array is $N=200$.
The distance from the atoms to the fiber surface is $r-a=200$ nm. The array period is $\Lambda=745$ nm, which satisfies the Bragg resonance condition for the field frequency $\omega_L=\omega_0$. Other parameters are as in Fig.~\ref{fig2}. 
}
\label{fig12}
\end{figure}

We now study the effect of the field detuning $\delta$ on the reflection and transmission of guided light by the atomic array.
We plot in Fig.~\ref{fig12} the powers $P_{f\xi}$ of the transmitted (left column) and reflected (middle column) quasilinearly polarized guided fields
as well as their total power $P_{\mathrm{tot}}$ (right column) as functions of the field detuning $\delta$. 
This figure is calculated for the array period $\Lambda=745$ nm, which satisfies the Bragg resonance condition for the field frequency $\omega_L=\omega_0$.
The left and right columns of the figure show that the power of the transmitted field as well as the total power of the guided fields have a local minimum at the position of the atomic resonance $\delta=0$. This behavior is opposite to the behavior around the Bragg resonance in the $\Lambda$ dependence (see the left and right columns of Fig.~\ref{fig11}).
The fact that $P_{\mathrm{tot}}$ has a peak at $\Lambda=745$ nm for $\delta=0$ and a minimum at $\delta=0$ for $\Lambda=745$ nm means that
the point $(\Lambda=745 \text{\ nm},\delta=0)$ is a saddle point in the two-dimensional profile
of $P_{\mathrm{tot}}$ with respect to the variables $\Lambda$ and $\delta$.
The occurrence of the local minimum of the total power of guided light indicates the occurrence
of the maximum of the power of light scattered into the radiation modes. 
Figure \ref{fig12}(e) shows that, when the input guided light is quasilinearly polarized along the minor principal axis $y$,
the scattering into the backward guided modes has a peak at the exact atomic resonance. This behavior is similar to that of the scattering into the radiation modes.
Meanwhile, we observe from Fig.~\ref{fig12}(b) that, when the input guided light is quasilinearly polarized along the major principal axis $x$,
the scattering into the backward guided modes has a double-peak structure, with
a flat-bottomed dip in the vicinity of the exact atomic resonance. The height of the plateau is $|R_\infty|^2\simeq0.087<1$, in agreement with Eq.~\eqref{x61}.

%%%%%%%%%%%%%%%%%%%%%%% Figure 12

\begin{figure}[tbh]
\begin{center}
  \includegraphics{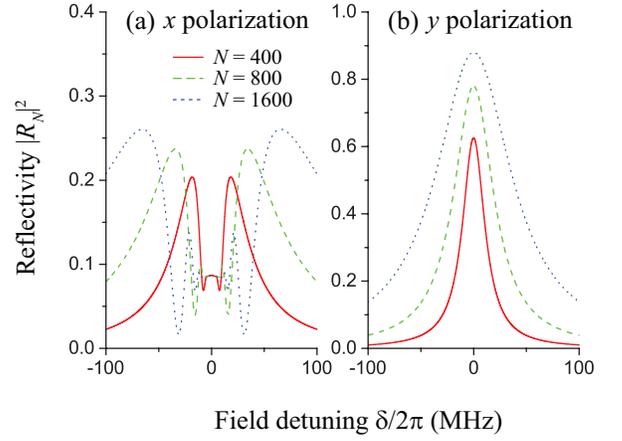}
 \end{center}
\caption{(Color online) Reflectivity $|R_N|^2$ of the atomic array as a function of the field detuning $\delta$ for three different values of the atom number $N$. 
The guided probe field is quasilinearly polarized along the major principal direction $x$ (left column) or 
the minor principal direction $y$ (right column). The number of atoms in the array is $N=400$ (solid red lines), 800 (dashed green lines), and 1600 (dotted blue lines).
The distance from the atoms to the fiber surface is $r-a=200$ nm. The array period is $\Lambda=745$ nm, which satisfies the Bragg resonance condition for the field frequency $\omega_L=\omega_0$.
Other parameters are as in Fig.~\ref{fig2}. 
}
\label{fig13}
\end{figure}

We plot in Fig.~\ref{fig13} the reflectivity $|R_N|^2$ of the atomic array 
as a function of the field detuning $\delta$ for three different values of the atom number $N$. 
The figure is calculated for the array period $\Lambda=745$ nm, which satisfies the Bragg resonance condition for the field frequency $\omega_L=\omega_0$.
Figure~\ref{fig13}(a) shows that, in the case of $x$-polarized guided fields, there is a plateau around the point $\delta=0$. 
The height of the plateau does not depend on the atom number $N$. However, the width of the plateau increases with increasing $N$.
The plateau is surrounded by two regions where modulations of $|R_N|^2$ occur. 
The maximum value of $|R_N|^2$ is achieved outside the plateau.
Figure~\ref{fig13}(b) shows that, in the case of $y$-polarized guided fields, $|R_N|^2$ has a central peak at $\delta=0$.
The linewidth of the peak increases with increasing $N$. In the limit $N\to\infty$, the reflectivity $|R_N|^2$ approaches unity.
The numerical results shown in Fig.~\ref{fig13} are in agreement with the analytical results of Sec. \ref{subsec:Bragg}.

%%%%%%%%%%%%%%%%%%%%%%% Figure 13

%%%%%%%%%%%%%%%%%%%%%%% Figure 14
\begin{figure}[tbh]
\begin{center}
  \includegraphics{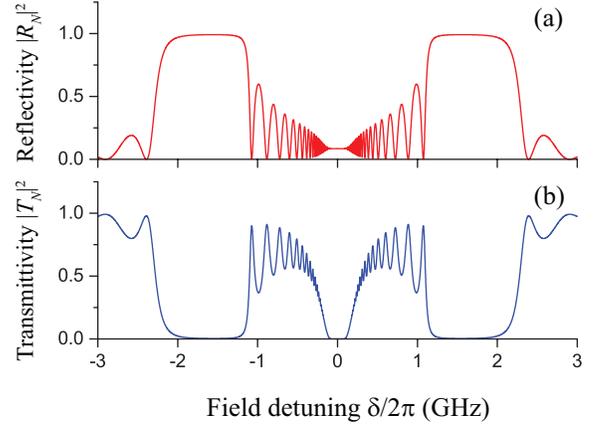}
 \end{center}
\caption{(Color online) Reflectivity $|R_N|^2$ (a) and transmittivity $|T_N|^2$ (b)  of the atomic array as  functions of the field detuning $\delta$ for the atom number $N=150,000$.
The guided probe field is quasilinearly polarized along the major principal direction $x$.
The distance from the atoms to the fiber surface is $r-a=200$ nm. The array period is $\Lambda=745$ nm. Other parameters are as in Fig.~\ref{fig2}. 
}
\label{fig14}
\end{figure}

%%%%%%%%%%%%%%%%%%%%%%% Figure 15
\begin{figure}[tbh]
\begin{center}
  \includegraphics{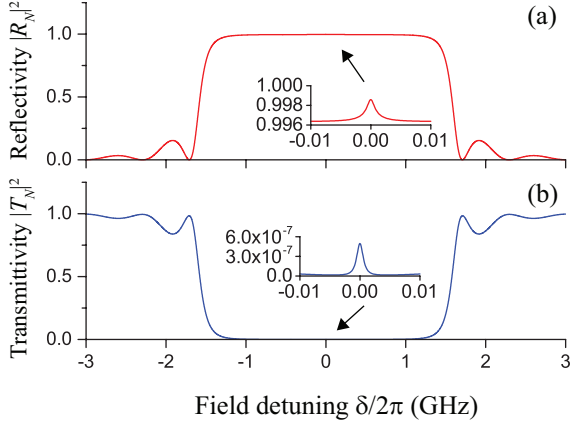}
 \end{center}
\caption{(Color online) Same as Fig.~\ref{fig14} but the guided probe field is quasilinearly polarized along the minor principal direction $y$. 
The insets show the narrow peak structures around the point $\delta=0$.
}
\label{fig15}
\end{figure}

%%%%%%%%%%%%%%%%%%%%%%% Figure 16
\begin{figure}[tbh]
\begin{center}
  \includegraphics{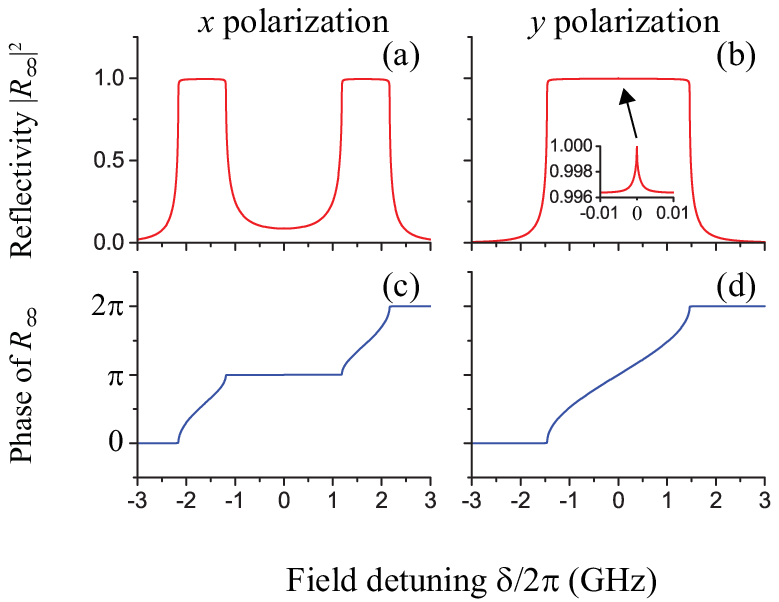}
 \end{center}
\caption{(Color online) Reflectivity $|R_\infty|^2$ (upper row) and phase $\varphi_{R_\infty}$ of the reflection coefficient $R_\infty$ (lower row) 
of an infinite atomic array as functions of the field detuning $\delta$.
The guided probe field is quasilinearly polarized along the major principal direction $x$ (left column) or the minor principal direction $y$ (right column).
The distance from the atoms to the fiber surface is $r-a=200$ nm. The array period is $\Lambda=745$ nm. Other parameters are as in Fig.~\ref{fig2}. 
}
\label{fig16}
\end{figure}

According to the previous section, in the neighborhood of the Bragg resonance, band gaps may be formed when the number of atoms $N$ in the array is much larger than 
the threshold atom number $N_{\mathrm{gap}}$. For the parameters of the nanofiber and the atoms used in our numerical calculations, we find $N_{\mathrm{gap}}\simeq 43,000$ and $33,000$ for the cases of $x$- and $y$-polarized guided light fields, respectively. In order to see the band gaps,
we plot in Figs.~\ref{fig14} and \ref{fig15} the reflectivity $|R_N|^2$ and the transmittivity $|T_N|^2$ of the atomic array 
as functions of the field detuning $\delta$ for a very large number of atoms, namely for $N=150,000$.
In addition, we plot in Fig.~\ref{fig16} the reflectivity $|R_\infty|^2$ (upper row) and  the phase of the reflection coefficient $R_\infty$ (lower row) of an infinite atomic array 
as functions of the field detuning $\delta$. 

Figure~\ref{fig14} shows that, in the case of $x$-polarized guided fields, in addition to a narrow plateau around the atomic resonance $\delta=0$, there are two wide plateaus, one on the left and one on the right. 
The left- and right-side plateaus correspond to the photonic band gaps. They extend from $-\Delta_{\mathrm{max}}$ to $-\Delta_{\mathrm{min}}$
and from $\Delta_{\mathrm{min}}$ to $\Delta_{\mathrm{max}}$, where $\Delta_{\mathrm{min}}$ and $\Delta_{\mathrm{max}}$ are given by  Eqs.~\eqref{x78}. 
We obtain  from Eqs.~\eqref{x78} the estimates $\Delta_{\mathrm{min}}=2\pi\times 1.19$ GHz
and $\Delta_{\mathrm{max}}=2\pi\times 2.16$ GHz for the parameters of Fig.~\ref{fig14}. 
In the band gap regions, we have $|R_N|^2\simeq1$ and $|T_N|^2\simeq0$. In the central plateau, we have 
$|R_N|^2\simeq0.087<1$, unlike in the band gaps, and $|T_N|^2\simeq0$, like in the band gaps. These estimates are in agreement with Eqs.~\eqref{x60} and \eqref{x61}.
The characteristic range of the central plateau is from $-\delta_{\mathrm{flat}}$
to $\delta_{\mathrm{flat}}$, where $\delta_{\mathrm{flat}}=\sqrt{\gamma_{s}\gamma N}/2 \simeq 2\pi\times 111$ MHz [see the discussion at the end of the paragraph around Eqs.~\eqref{x60} and \eqref{x61}].

Figure~\ref{fig15} shows that, in the case of $y$-polarized guided fields, 
there is a wide plateau around the atomic resonance $\delta=0$. 
This plateau corresponds to the set of the two photonic band gaps that 
extend from $-\Delta_{\mathrm{max}}$ to 0 and from 0 to $\Delta_{\mathrm{max}}$, where $\Delta_{\mathrm{max}}$ is given by Eq.~\eqref{x85}. We obtain  from Eq.~\eqref{x85} the estimate $\Delta_{\mathrm{max}}=2\pi\times 1.46$ GHz for the parameters of Fig.~\ref{fig15}. 
In the band gap region, we have $|R_N|^2\simeq1$ and $|T_N|^2\simeq0$.
The insets of Figs.~\ref{fig15}(a) and \ref{fig15}(b) show that there is a narrow and low peak
at the atomic resonance $\delta=0$, in agreement with the result of Chang \cite{Chang14}.
Since $\delta_{\mathrm{lat}}=0$ in the case of Fig.~\ref{fig15}, the atomic resonance $\delta=0$ is the common edge of the two band gaps. 
When $\delta_{\mathrm{lat}}\not=0$, the region of detunings from $0$ to $\delta_{\mathrm{lat}}$ is excluded from the band gap and we may observe clearly the separation between the two band gaps \cite{Deutsch95}. 

Figures~\ref{fig16}(a) and \ref{fig16}(b) show the reflectivity $|R_\infty|^2$ of the array in the case of an infinite large number of atoms. In this limit, we have $|T_\infty|^2=0$. We observe from Figs.~\ref{fig16}(c) and \ref{fig16}(d) that the phase $\varphi_{R_\infty}$ of the reflection coefficient $R_\infty$
increases with increasing field detuning $\delta$, that is, with increasing field frequency $\omega_L$, in the band gap regions. The positive slope of the frequency dependence of the phase $\varphi_{R_\infty}$ indicates the group delay of the reflected guided field in the band gap regions. The change of the phase $\varphi_{R_\infty}$ over the band gap is equal to $\pi$ for the gap width of about $2\pi\times 1$ GHz in the case of Fig.~\ref{fig16}(c) and is equal to $2\pi$ for 
the gap width of about $2\pi\times 3$ GHz in the case of Fig.~\ref{fig16}(d). 
Hence, the group delay $\tau_{\mathrm{delay}}=\varphi'_{R_\infty}(\omega)$ is estimated to be about $0.5$ and $0.3$ ns in the cases of $x$- and $y$-polarized guided fields, respectively.

We note that the numerical results for the band gaps presented in Figs.~\ref{fig14} and \ref{fig15} are of just academic interest.
One reason is that the formation of the band gaps requires a very large number of atoms trapped in a linear array along the fiber.
Another reason is that the band gaps extend over a large region of frequency where, in addition to the hyperfine level $6P_{3/2}F'=5$ considered in this paper,
the hyperfine levels $6P_{3/2}F'=4$ and $6P_{3/2}F'=3$ must also be accounted for. 
However, we expect that the qualitative aspects of the band gap formation remain valid under the condition of large detunings.

\section{Summary}
\label{sec:summary}

We have studied the propagation of guided light in an array of multilevel cesium atoms with the transitions between the hyperfine levels $6S_{1/2}F=4$ and $6P_{3/2}F'=5$ of the $D_2$ line outside a nanofiber.
We have derived the coupled-mode propagation equation, the input-output equation, the scattering matrix, and the transfer matrix in the linear, quasistationary, weak-excitation regime.
We have taken into account the complexity of the polarization of the guided field, the multilevel structure of the atoms, and the discreteness of the atomic positions in the array.  
The general solution of the input-output equation has been obtained. 
We have found that, when the initial distribution of populations of atomic ground-state sublevels is flat, the quasilinear polarizations along the principal axes $x$ and $y$, which are parallel and perpendicular, respectively, 
to the radial direction of the atomic position, are not coupled to each other in the linear coherent scattering process.
Reflection and transmission amplitudes have been calculated and analyzed in a variety of cases.
We found that, when the array period is far from the Bragg resonance, the 
backward scattering is very weak and modulates with increasing number of atoms in the array. Such oscillations are the results of  the interference between the light waves reflected from different atoms in the array. The modulation period is determined by the mismatch between the array period $\Lambda$ and the in-fiber half-wavelength $\lambda_F/2=\pi/\beta_L$ of the probe light.
In the neighborhood of the Bragg resonance, most of the guided light can be reflected back in a broad region of field detunings even though there is an irreversible decay channel into radiation modes.
Due to the collective effect and the Bragg resonance condition, the loss due to emission into radiation modes can be suppressed.
When the input guided light is quasilinearly polarized along the major principal axis $x$, under the Bragg resonance condition, 
the frequency dependence of the reflectivity of the atomic array has a double-peak structure, with
a flat-bottomed (plateau-shaped) dip in the vicinity of the exact atomic resonance. 
The value of the reflectivity of this central plateau area does not depend on the number of atoms, the field detuning, and the dipole
matrix elements, and is determined just by the ratio between the longitudinal and radial components of the guided-mode profile function. 
When the input guided light is quasilinearly polarized along the minor principal axis $y$, under the Bragg resonance condition,
the frequency dependence of the reflectivity of the atomic array has a broad peak at the atomic resonance. 
When the atom number is large enough, two different band gaps may be formed.
When the input guided light is quasilinearly polarized along the major principal axis $x$, the two band gaps are well separated from each other.
However, when the input guided light is quasilinearly polarized along the minor principal axis $y$, the two band gaps are, in general, close to each other, and have a common edge
when the atomic resonance coincides with the Bragg resonance.

We emphasize that incoherent scattering into guided modes is not accounted for in our formalism. Because of this limitation, our results are valid only in the cases where coherent scattering is dominant to incoherent scattering. When the array period is not far away from the Bragg resonance condition, due to the collective enhancement, coherent scattering is significant.
Our results are meaningful in this case.

\begin{acknowledgments}
We thank C. Clausen, C. Sayrin, and P. Schneeweiss for helpful comments and discussions.
F.L.K. acknowledges support by the Austrian Science Fund (Lise Meitner Project No. M 1501-N27)
and by the European Commission (Marie Curie IIF Grant No. 332255). 
\end{acknowledgments}

%%%%%%%%%%%%%

\appendix

\section{Guided modes of a nanofiber}
\label{sec:guided}

Consider a nanofiber that is a silica cylinder of radius $a$ and refractive index $n_1$ and is surrounded by an infinite background medium of refractive index $n_2$,
where $n_2<n_1$. The radius of the nanofiber is well below a given free-space wavelength $\lambda$ of light. Therefore, the nanofiber supports only the hybrid fundamental modes HE$_{11}$ corresponding to the given wavelength $\lambda$ \cite{fiber books}. For a fundamental guided mode HE$_{11}$ of a light field of frequency $\omega$ (free-space wavelength $\lambda=2\pi c/\omega$ and free-space wave number $k=\omega/c$), the propagation constant $\beta$ is determined by the
fiber eigenvalue equation \cite{fiber books}
\begin{eqnarray}\label{g1}
\frac{J_0(h a)}{h a J_1(h a)}&=&
-\frac{n_1^2+n_2^2}{2n_1^2}\frac{K_1'(q a)}{q a K_1(q a)}+ \frac{1}{h^2 a^2}
\nonumber\\&&\mbox{}
-\Bigg[\left(\frac{n_1^2-n_2^2}{2n_1^2}\frac{K_1'(q a)}{q a K_1(q a)}\right)^2
\nonumber\\&&\mbox{}
+\frac{\beta^2}{n_1^2 k^2}\left(\frac{1}{q^2a^2}+\frac{1}{h^2a^2}\right)^2 \Bigg]^{1/2}.
\end{eqnarray}
Here,  the parameters $h=(n_1^2k^2-\beta^2)^{1/2}$ and $q=(\beta^2-n_2^2k^2)^{1/2}$ characterize the fields inside and outside the fiber, respectively. The notations $J_n$ and $K_n$ stand for the Bessel functions of the first kind and the modified Bessel functions of the second kind, respectively. 

According to \cite{fiber books}, the cylindrical-coordinate vector components of the profile function $\mathbf{e}(\mathbf{r})$ 
of the electric part of the fundamental guided mode that propagates in the forward ($+\hat{\mathbf{z}}$) direction and is
counterclockwise quasicircularly polarized are given, for $r<a$, by
\begin{eqnarray}\label{g2}
e_{r}&=&iC\frac{q}{h}\frac{K_1(qa)}{J_1(ha)}[(1-s)J_0(hr)-(1+s)J_2(hr) ],
\nonumber\\
e_{\varphi}&=&-C\frac{q}{h}\frac{K_1(qa)}{J_1(ha)}[(1-s)J_0(hr)+(1+s)J_2(hr) ],
\nonumber\\
e_{z}&=&C\frac{2q}{\beta}\frac{K_1(qa)}{J_1(ha)}J_1(hr),
\end{eqnarray}
and, for $r>a$, by
\begin{eqnarray}\label{g3}
e_{r}&=&iC[(1-s)K_0(qr)+(1+s)K_2(qr) ],
\nonumber\\
e_{\varphi}&=&-C[(1-s)K_0(qr)-(1+s)K_2(qr) ],
\nonumber\\
e_{z}&=&C\frac{2q}{\beta}K_1(qr).
\end{eqnarray}
Here,  the parameter $s$ is defined as
\begin{equation}\label{g4} 
s=\frac{{1}/{h^2a^2}+{1}/{q^2a^2}}{{J_1^\prime (ha)}/{haJ_1(ha)}+{K_1^\prime (qa)}/{qaK_1(qa)}}.
\end{equation}
The parameter $C$ is the normalization coefficient. We take $C$ to be a positive real number and use the normalization condition 
\begin{equation}\label{g5}
\int _{0}^{2\pi}d\varphi\int _{0}^{\infty}n_{\mathrm{ref}}^2\,|\mathbf{e}|^2r\,dr=1.
\end{equation}
Here,  $n_{\mathrm{ref}}(r)=n_1$ for $r<a$, and $n_{\mathrm{ref}}(r)=n_2$ for $r>a$.

We label quasicircularly polarized fundamental guided modes HE$_{11}$ by using a mode index $\mu=(\omega,f,l)$, where $\omega$ is the mode frequency, $f=+1$ or $-1$ (or simply $+$ or $-$) 
denotes the forward ($+\hat{\mathbf{z}}$) or backward ($-\hat{\mathbf{z}}$) propagation direction, respectively, and $l=+1$ or $-1$ (or simply $+$ or $-$) 
denotes the counterclockwise  or clockwise circulation, respectively, of the transverse component of the polarization around the axis $+\hat{\mathbf{z}}$. 
In the cylindrical coordinates, the components of the profile function $\mathbf{e}^{(\mu)}(\mathbf{r})$ of the electric part of the quasicircularly polarized fundamental guided mode $\mu$ are given by
\begin{eqnarray}\label{g6}
e_{r}^{(\mu)}&=&e_{r},
\nonumber\\
e_{\varphi}^{(\mu)}&=&le_{\varphi},
\nonumber\\
e_{z}^{(\mu)}&=& fe_{z}.
\end{eqnarray}
Consequently, the profile function of the quasicircularly polarized mode $(\omega, f, l)$ can be written as
\begin{equation}\label{g7}
\mathbf{e}^{(\omega fl)}=\hat{\mathbf{r}}e_r+l\hat{\boldsymbol{\varphi}}e_\varphi+f\hat{\mathbf{z}}e_z,
\end{equation}
where the notations 
$\hat{\mathbf{r}} = \hat{\mathbf{x}}\cos\varphi + \hat{\mathbf{y}}\sin\varphi$,  
$\hat{\boldsymbol{\varphi}} = -\hat{\mathbf{x}}\sin\varphi + \hat{\mathbf{y}}\cos\varphi$, 
and $\hat{\mathbf{z}}$ stand for the unit basis vectors of the cylindrical coordinate system $\{r,\varphi,z\}$.
Here,  $\hat{\mathbf{x}}$ and $\hat{\mathbf{y}}$ are the unit basis vectors of the Cartesian coordinate system for the fiber transverse plane $xy$.

We introduce the notations $V_0=V_z$ and $V_{\pm 1}=\mp(V_x\pm i V_y)/\sqrt{2}$ for the spherical tensor components of an arbitrary vector $\mathbf{V}$. 
Due to the properties of the guided-mode profile functions \cite{fiber books}, we can represent
the spherical tensor components $e_{q}^{(\omega fl)}$ of the profile function $\mathbf{e}^{(\omega fl)}$ of the quasicircularly polarized guided mode $(\omega, f, l)$ in the form
\begin{equation}\label{g10}
e_{q}^{(\omega fl)}=f^{1+q}e^{iq(\varphi-\pi/2)}|e_{ql}|.
\end{equation}
Here,  we have introduced the notations
\begin{eqnarray}\label{g11} 
|e_0|&=&|e_z|,\nonumber\\
|e_{+1}|&=&\frac{|e_r|-|e_{\varphi}|}{\sqrt{2}},\nonumber\\ 
|e_{-1}|&=&\frac{|e_r|+|e_{\varphi}|}{\sqrt{2}}. 
\end{eqnarray}

We note that expression \eqref{g7} for the mode profile function $\mathbf{e}^{(\omega fl)}$ does not include the phase factor $e^{if\beta z +il\varphi}$,
which is present in the full expression for the electric part of the guided field in a quasicircularly polarized mode.
Indeed, the electric part $\boldsymbol{\mathcal{E}}_{\mathrm{circ}}^{(\omega fl)}$ of the guided field in the quasicircularly polarized mode $(\omega, f, l)$
is given by \cite{fiber books}
\begin{equation}\label{g7a}
\boldsymbol{\mathcal{E}}_{\mathrm{circ}}^{(\omega fl)} = A(\hat{\mathbf{r}}e_r+l\hat{\boldsymbol{\varphi}}e_\varphi+
f\hat{\mathbf{z}}e_z) e^{if\beta z +il\varphi},
\end{equation}
where the coefficient $A$ is determined by the power of the field.

Quasilinearly polarized guided modes are linear superpositions of quasicircularly polarized guided modes.
The profile functions of quasilinearly polarized modes $(\omega,f,\xi)$, where $\xi=x$ or $y$, are given by
\begin{eqnarray}\label{g9}
\mathbf{e}^{(\omega fx)}&=&\sqrt2(\hat{\mathbf{r}}e_r\cos\varphi+i\hat{\boldsymbol{\varphi}}e_\varphi\sin\varphi+f\hat{\mathbf{z}}e_z\cos\varphi),\nonumber\\
\mathbf{e}^{(\omega fy)}&=&\sqrt2(\hat{\mathbf{r}}e_r\sin\varphi-i\hat{\boldsymbol{\varphi}}e_\varphi\cos\varphi+f\hat{\mathbf{z}}e_z\sin\varphi).\qquad
\end{eqnarray}
The electric part $\boldsymbol{\mathcal{E}}_{\mathrm{lin}}^{(\omega f\xi)}$ of the guided field in a quasilinearly polarized mode $(\omega, f, \xi)$ is given by \cite{fiber books}
\begin{equation}\label{g7d}
\boldsymbol{\mathcal{E}}_{\mathrm{lin}}^{(\omega f\xi)} 
=A\mathbf{e}^{(\omega f\xi)}e^{if\beta z}.
\end{equation}

We now examine the coefficients of spontaneous emission from a multilevel atom in the vicinity of a nanofiber into the guided modes. We use the notations $|e\rangle$ and $|g\rangle$ for the magnetic sublevels of the atom. 
The coefficients 
\begin{eqnarray}\label{g13}
G_{\omega fl eg}=\sqrt{\frac{\omega\beta'}{4\pi\epsilon_0\hbar}}\;
\big(\mathbf{d}_{eg}\cdot\mathbf{e}^{(\omega fl)}\big)e^{i(f\beta z+l\varphi)}
\end{eqnarray}
with $l=+$ or $-$ characterize the coupling of the atomic transitions $|e\rangle\leftrightarrow |g\rangle$ with the quasicircularly polarized guided modes $(\omega, f, l)$, and the coefficients
\begin{eqnarray}\label{g14}
G_{\omega f\xi eg}=\sqrt{\frac{\omega\beta'}{4\pi\epsilon_0\hbar}}\;
\big(\mathbf{d}_{eg}\cdot\mathbf{e}^{(\omega f\xi)}\big)e^{if\beta z}
\end{eqnarray}
with $\xi=x$ or $y$ characterize the coupling of the atomic transitions $|e\rangle\leftrightarrow |g\rangle$ with the quasilinearly polarized guided modes $(\omega, f, \xi)$.
Here,  the notation $\beta'$ stands for the derivative of the propagation constant $\beta$ with respect to the frequency $\omega$, and the notation $\mathbf{d}_{eg}$ stands for the atomic dipole matrix element.

According to Ref.~\cite{cesium decay}, the spontaneous emission from the atom into the guided modes of the nanofiber affects the evolution of the reduced density matrix of the atom through the set of decay coefficients 
\begin{eqnarray}\label{g21}
\gamma_{ee'gg'}^{(\mathrm{gyd})}&=&\sum_{f=+,-}\gamma_{ee'gg'}^{(f)},\nonumber\\
\gamma_{ee'}^{(\mathrm{gyd})}&=&\sum_{f=+,-}\gamma_{ee'}^{(f)}.
\end{eqnarray}
Here,  we have introduced the notations
\begin{eqnarray}\label{g20}
\gamma_{ee'gg'}^{(f)}&=&\sum_{l=+,-}\gamma_{ee'gg'}^{(fl)}=\sum_{\xi=x,y}\gamma_{ee'gg'}^{(f\xi)},\nonumber\\
\gamma_{ee'}^{(f)}&=&\sum_{l=+,-}\gamma_{ee'}^{(fl)}=\sum_{\xi=x,y}\gamma_{ee'}^{(f\xi)}, 
\end{eqnarray}
where
\begin{eqnarray}\label{g18}
\gamma_{ee'gg'}^{(fp)}&=&\gamma_{ee'gg'}^{(fp fp)},\nonumber\\
\gamma_{ee'}^{(fp)}&=&\gamma_{ee'}^{(fp fp)},
\end{eqnarray}
with
\begin{subequations}\label{g19}
\begin{eqnarray}
\gamma_{ee'gg'}^{(fp f'p')}&=&2\pi G_{\omega_0fp eg}G_{\omega_0f'p' e'g'}^*,\label{g19a}\\
\gamma_{ee'}^{(fp f'p')}&=&\sum_g\gamma_{ee'gg}^{(fp f'p')}.\label{g19b}
\end{eqnarray}
\end{subequations}

According to Eq.~(\ref{x3}), only one spherical tensor component $d_{eg}^{(q)}\equiv (d_{eg})_q$ of the dipole vector $\mathbf{d}_{eg}$, with $q=M_e-M_g=-1$, 0, or 1, is nonzero. 
Hence, we obtain the formula
\begin{equation}\label{g22}
G_{\omega fl eg}=f^{1+q}e^{-iq\pi/2}e^{if\beta z}e^{i(l-q)\varphi}\sqrt{\frac{\omega\beta'}{4\pi\epsilon_0\hbar}}
d_{eg}^{(q)} |e_{-ql}|.
\end{equation}
From Eq.~(\ref{g10}), we find
\begin{equation}\label{g30}
e_{q}^{(\omega fl)}=(-1)^q   e^{2iq\varphi} e_{-q}^{(\omega f\bar{l})},
\end{equation}
where $\bar{l}=-l$. On the other hand, when we use the properties of the Clebsch-Gordan coefficients and Eq.~(\ref{x3}), we find 
\begin{equation}\label{g31}
d_{eg}^{(q)}=(-1)^{F-F'+1} d_{\bar{e}\bar{g}}^{(\bar{q})},
\end{equation}
where $\bar{e}$ and $\bar{g}$ are the levels $|F',-M_e\rangle$ and $|F,-M_g\rangle$, respectively,
$q=M_e-M_g$, and $\bar{q}=-q$.
Then, we obtain the relation
\begin{equation}\label{g32}
G_{\omega fleg}=(-1)^{F-F'+1+M_e-M_g} e^{-2i(M_e-M_g-l)\varphi}G_{\omega f\bar{l}\bar{e}\bar{g}},
\end{equation}
which leads to
\begin{equation}\label{g33}
G_{\omega fleg}G_{\omega f'l'eg}^*=e^{2i(l-l')\varphi}G_{\omega f\bar{l}\bar{e}\bar{g}}G_{\omega f'\bar{l'}\bar{e}\bar{g}}^*.
\end{equation}

\section{Radiation modes of a nanofiber}
\label{sec:radiation}

For the radiation modes, we have $-kn_2<\beta<kn_2$.
The characteristic parameters for the field in the inside and outside of the fiber are $h=\sqrt{k^2n_1^2-\beta^2}$ and $q=\sqrt{k^2n_2^2-\beta^2}$, respectively.
The mode functions of the electric parts of the radiation modes $\nu=(\omega\beta m l)$
\cite{fiber books} are given, for $r<a$, by
\begin{eqnarray}\label{q1}
e_r^{(\nu)}&=&
\frac{i}{h^2}\left[\beta hAJ'_m(hr)+im\frac{\omega\mu_0}{r}BJ_m(hr)\right],\nonumber\\ 
e_{\varphi}^{(\nu)}&=&
\frac{i}{h^2}\left[im\frac{\beta}{r}AJ_m(hr)-h\omega\mu_0BJ'_m(hr)\right],\nonumber\\
e_z^{(\nu)}&=&AJ_m(hr),
\end{eqnarray}
and, for $r>a$, by 
\begin{eqnarray}\label{q2}
e_r^{(\nu)}&=&
\frac{i}{q^2}\sum_{j=1,2}
\left[\beta q C_jH^{(j)\prime}_m(qr)+im\frac{\omega\mu_0}{r}D_jH^{(j)}_m(qr)\right],\nonumber\\
e_{\varphi}^{(\nu)}&=&
\frac{i}{q^2}\sum_{j=1,2}
\left[im\frac{\beta}{r}C_jH^{(j)}_m(qr)-q\omega\mu_0D_jH^{(j)\prime}_m(qr)\right], \nonumber\\
e_z^{(\nu)}&=&\sum_{j=1,2}C_jH_m^{(j)}(qr).
\end{eqnarray}
Here, $H_m^{(j)}$ with $m=0,\pm1,\pm2,\dots$ and $j=1,2$ is the Hankel function of the $m$-th order and the $j$-th kind, and 
$A$ and $B$ as well as $C_j$ and $D_j$ with $j=1,2$ are coefficients.
The coefficients $C_j$ and $D_j$ are related to the coefficients $A$ and $B$ as 
\cite{Tromborg}
\begin{eqnarray}\label{q3}
C_j&=&(-1)^{j}\frac{i\pi q^2a}{4n_2^2}(AL_j+i\mu_0cBV_j),\nonumber\\
D_j&=&(-1)^{j-1}\frac{i\pi q^2a}{4}(i\epsilon_0cAV_j-BM_j),
\end{eqnarray}
where
\begin{eqnarray}\label{q4}
V_j&=&\frac{mk\beta}{ah^2q^2}
(n_2^2-n_1^2)
J_m(ha)H_m^{(j)*}(qa),\nonumber\\
M_j&=&\frac{1}{h}J'_m(ha)H_m^{(j)*}(qa)
-\frac{1}{q}J_m(ha)H_m^{(j)*\prime}(qa),\nonumber\\
L_j&=&\frac{n_1^2}{h}J'_m(ha)H_m^{(j)*}(qa)
-\frac{n_2^2}{q}J_m(ha)H_m^{(j)*\prime}(qa).\nonumber\\
\end{eqnarray}
We specify two polarizations by choosing $B=i\eta A$ and $B=-i\eta A$ for $l=+$
and $l=-$, respectively. We take $A$ to be a real number
The orthogonality of the modes requires
\begin{eqnarray}\label{q5}
&&\int _0^{2\pi}d\varphi\int _{0}^{\infty}n_{\mathrm{ref}}^2
\left[\mathbf{e}^{(\nu)}\mathbf{e}^{(\nu')*}\right]_{\beta=\beta',m=m'}
\;rdr \nonumber\\&&
=N_{\nu}\delta_{ll'}\delta(\omega-\omega').
\end{eqnarray}
This leads to
\begin{equation}\label{q6}
\eta=\epsilon_0c\sqrt{\frac{n_2^2|V_j|^2+|L_j|^2}{|V_j|^2+n_2^2|M_j|^2}}.
\end{equation}
The constant $N_{\nu}$ is given by 
\begin{equation}\label{q7}
N_{\nu}=\frac{8\pi \omega}{q^2}\left(n_2^2|C_j|^2+\frac{\mu_0}{\epsilon_0}|D_j|^2\right).
\end{equation}
We use the normalization $N_{\nu}=1$. 

We now examine the coefficients of spontaneous emission from a multilevel atom in the vicinity of a nanofiber into the radiation modes. 
We use the notations $|e\rangle$ and $|g\rangle$ for the magnetic sublevels of a multilevel atom in the vicinity of the nanofiber.
According to Ref.~\cite{cesium decay}, the spontaneous emission from the atom into the radiation modes of the nanofiber affects the evolution of the atomic reduced density matrix through the set of decay coefficients 
 \begin{eqnarray}\label{q14}
\gamma^{(\mathrm{rad})}_{ee'gg'}&=&2\pi \sum_{ml}\int_{-k_0n_2}^{k_0n_2}d\beta\,G_{\nu_0 eg}G_{\nu_0 e'g'}^*,
\nonumber\\
\gamma^{(\mathrm{rad})}_{ee'}&=&2\pi \sum_{mlg}\int_{-k_0n_2}^{k_0n_2}d\beta\, G_{\nu_0 eg}G_{\nu_0 e'g}^*.
\end{eqnarray}
Here,  $\nu_0=(\omega_0,\beta,m,l)$ labels resonant radiation modes and
\begin{equation}\label{q15}
G_{\nu eg}=\sqrt{\frac{\omega}{4\pi\epsilon_0\hbar}}\;
\big(\mathbf{d}_{eg}\cdot\mathbf{e}^{(\nu)}\big)e^{i(\beta z+m\varphi)}
\end{equation}
characterizes the coupling of the atomic transition $|e\rangle\leftrightarrow |g\rangle$ with
the radiation mode $\nu=(\omega,\beta,m,l)$. 

\section{Properties of the scattering matrix}
\label{sec:properties}

In this appendix, we discuss the properties of the single-atom guided-field scattering matrix $\mathbf{S}$, given by Eq.~\eqref{x31}. 
For this purpose, we calculate the matrix elements $S_{f\xi f'\xi'}$ of $\mathbf{S}$ in the mode basis formed by the quasilinearly polarized modes with the indices $f=\pm$ and $\xi=x,y$. 
We assume that the atom is positioned on the $x$ axis, that is, the coordinates of the atomic array in the fiber transverse plane are $(x=x_0,y=0)$. 
We consider the case where the initial population distribution $p_g$ is flat. 
With an appropriate choice of the phase of the mode functions, the coupling coefficients $\mathcal{G}_{\omega f\xi eg}$
in the mode polarization basis $\xi=x,y$ can be expressed in terms of the coupling coefficients $\mathcal{G}_{\omega fl eg}$
in the mode polarization basis $l=+,-$ as
\begin{eqnarray}\label{v1}
\mathcal{G}_{\omega fx eg}&=&\frac{1}{\sqrt2}\big(\mathcal{G}_{\omega f+ eg}+\mathcal{G}_{\omega f- eg}\big),
\nonumber\\
\mathcal{G}_{\omega fy eg}&=&\frac{1}{i\sqrt2}\big(\mathcal{G}_{\omega f+ eg}-\mathcal{G}_{\omega f- eg}\big).
\end{eqnarray}
Then, Eq.~(\ref{x31}) yields
\begin{eqnarray}\label{v2}
S_{fxf'x}&=&\frac{f}{(\gamma-2i\delta)(2F+1)}\sum_{eg}
(\mathcal{G}_{f+eg}^*\mathcal{G}_{f'+eg}
\nonumber\\&&\mbox{}
+\mathcal{G}_{f-eg}^*\mathcal{G}_{f'-eg}+\mathcal{G}_{f+eg}^*\mathcal{G}_{f'-eg}+\mathcal{G}_{f-eg}^*\mathcal{G}_{f'+eg}),
\nonumber\\
S_{fyf'y}&=&\frac{f}{(\gamma-2i\delta)(2F+1)}\sum_{eg}
(\mathcal{G}_{f+eg}^*\mathcal{G}_{f'+eg}
\nonumber\\&&\mbox{}
+\mathcal{G}_{f-eg}^*\mathcal{G}_{f'-eg}-\mathcal{G}_{f+eg}^*\mathcal{G}_{f'-eg}-\mathcal{G}_{f-eg}^*\mathcal{G}_{f'+eg}),
\nonumber\\
S_{fxf'y}&=&\frac{f}{i(\gamma-2i\delta)(2F+1)}\sum_{eg}
(\mathcal{G}_{f+eg}^*\mathcal{G}_{f'+eg}
\nonumber\\&&\mbox{}
-\mathcal{G}_{f-eg}^*\mathcal{G}_{f'-eg}-\mathcal{G}_{f+eg}^*\mathcal{G}_{f'-eg}+\mathcal{G}_{f-eg}^*\mathcal{G}_{f'+eg}),
\nonumber\\
S_{fyf'x}&=&-\frac{f}{i(\gamma-2i\delta)(2F+1)}\sum_{eg}
(\mathcal{G}_{f+eg}^*\mathcal{G}_{f'+eg}
\nonumber\\&&\mbox{}
-\mathcal{G}_{f-eg}^*\mathcal{G}_{f'-eg}+\mathcal{G}_{f+eg}^*\mathcal{G}_{f'-eg}-\mathcal{G}_{f-eg}^*\mathcal{G}_{f'+eg}),
\nonumber\\
\end{eqnarray}
where $\mathcal{G}_{fl eg}=\mathcal{G}_{\omega_L fl eg}$.

We note that the $z$-independent coupling coefficients $\mathcal{G}_{\omega fpeg}$ and the $z$-dependent coupling coefficients $G_{\omega fpeg}$ are related to each other via the formula
\begin{equation}\label{v3}
G_{\omega fpeg}=\frac{1}{\sqrt{2\pi}}\mathcal{G}_{\omega fpeg} e^{if\beta z}.
\end{equation}
In terms of the $z$-independent coupling coefficients $\mathcal{G}_{\omega fpeg}$, Eq.~\eqref{g19a} for the coefficients $\gamma_{ee'gg'}^{(fp f'p')}$ of spontaneous emission into guided modes can be rewritten as
\begin{equation}\label{v3a}
\gamma_{ee'gg'}^{(fp f'p')}=\mathcal{G}_{\omega_0fp eg}\mathcal{G}_{\omega_0f'p' e'g'}^* e^{i(f-f')\beta_0 z}.
\end{equation}
For an individual atomic transition $|e\rangle \to |g\rangle$ and an individual guided mode $fp$, the rate of spontaneous emission is given by
\begin{equation}\label{v3b}
\gamma_{eg}^{(fp)}=\mathcal{G}_{\omega_0fp eg}\mathcal{G}_{\omega_0fp eg}^*.
\end{equation}

We insert expression \eqref{v3} into Eq.~\eqref{g33} and apply the summation over $eg$ to the result. In addition, we take the azimuthal angle $\varphi=0$, 
which corresponds to the atom on the axis $x$.
Then, we find 
\begin{eqnarray}\label{v4}
\sum_{eg}\mathcal{G}_{fleg}^*\mathcal{G}_{f'l'eg}&=&\sum_{eg}\mathcal{G}_{f\bar{l}eg}^*\mathcal{G}_{f'\bar{l'}eg}.
\end{eqnarray}
Due to the above property, we find from Eqs.~\eqref{v2} the relation
\begin{equation}\label{v5} 
S_{fxf'y}=S_{fyf'x}=0. 
\end{equation}
Thus, the quasilinear polarizations along the principal axes $x$ and $y$, which are parallel and perpendicular, respectively, 
to the radial direction of the atomic position, are not coupled to each other in the linear coherent scattering process.

In addition to Eq.~\eqref{v5}, we find from Eqs.~\eqref{v2} the following expressions for the nonzero matrix elements $S_{fxf'x}$ and $S_{fyf'y}$ of the scattering matrix:
\begin{eqnarray}\label{v6}
S_{fxf'x}&=&\frac{2f}{(\gamma-2i\delta)(2F+1)}\sum_{eg}
(\mathcal{G}_{f+eg}^*\mathcal{G}_{f'+eg}
\nonumber\\&&\mbox{}
+\mathcal{G}_{f+eg}^*\mathcal{G}_{f'-eg}),
\nonumber\\
S_{fyf'y}&=&\frac{2f}{(\gamma-2i\delta)(2F+1)}\sum_{eg}
(\mathcal{G}_{f+eg}^*\mathcal{G}_{f'+eg}
\nonumber\\&&\mbox{}
-\mathcal{G}_{f+eg}^*\mathcal{G}_{f'-eg}).
\end{eqnarray}

When we insert Eq.~(\ref{v3}) into Eq.~(\ref{g22}), we obtain the explicit expression 
\begin{equation}\label{v7}
\mathcal{G}_{fleg}=f^{1+q}e^{-iq\pi/2}\sqrt{\frac{\omega_L}{2\epsilon_0\hbar v_g}}
d_{eg}^{(q)} |e_{-ql}|,
\end{equation}
which yields
\begin{equation}\label{v8}
\mathcal{G}_{fleg}^*\mathcal{G}_{f'l'eg}
=(ff')^{1+q}\frac{\omega_L}{2\epsilon_0\hbar v_g}|d_{eg}^{(q)}|^2 |e_{-ql}||e_{-ql'}|.
\end{equation}
Inserting expression (\ref{v8}) into Eqs.~(\ref{v6}), we find
\begin{eqnarray}\label{v9}
S_{fxf'x}&=&\frac{f}{(\gamma-2i\delta)(2F+1)}\frac{\omega_L}{\epsilon_0\hbar v_g}
\nonumber\\&&\mbox{}
\times\sum_{egq}(ff')^{1+q}|d_{eg}^{(q)}|^2 (|e_{-q}|^2+|e_{-q}||e_{q}|),
\nonumber\\
S_{fyf'y}&=&\frac{f}{(\gamma-2i\delta)(2F+1)}\frac{\omega_L}{\epsilon_0\hbar v_g}
\nonumber\\&&\mbox{}
\times\sum_{egq}(ff')^{1+q}|d_{eg}^{(q)}|^2 (|e_{-q}|^2-|e_{-q}||e_{q}|).\qquad
\end{eqnarray}
Furthermore, with the help of Eqs.~(\ref{g11}), we can show that 
\begin{equation}\label{v10}
|e_{-q}|^2+|e_{-q}||e_{q}|=
\begin{cases}
2|e_z|^2 & \text{if } q=0,\\
|e_r|^2+|e_r||e_\varphi| & \text{if } q=1,\\
|e_r|^2-|e_r||e_\varphi| & \text{if } q=-1,
\end{cases}
\end{equation}
and
\begin{equation}\label{v11}
|e_{-q}|^2-|e_{-q}||e_{q}|=
\begin{cases}
0 & \text{if } q=0,\\
|e_\varphi|^2+|e_r||e_\varphi| & \text{if } q=1,\\
|e_\varphi|^2-|e_r||e_\varphi| & \text{if } q=-1.
\end{cases}
\end{equation}
Hence, we find
\begin{eqnarray}\label{v12}
S_{fxf'x}&=&\frac{f}{(\gamma-2i\delta)(2F+1)}\frac{\omega_L}{\epsilon_0\hbar v_g}
\sum_{eg}\Big[2ff'|d_{eg}^{(0)}|^2 |e_z|^2 
\nonumber\\&&\mbox{}
+\big(|d_{eg}^{(1)}|^2+|d_{eg}^{(-1)}|^2\big) |e_r|^2\Big],
\nonumber\\
S_{fyf'y}&=&\frac{f}{(\gamma-2i\delta)(2F+1)}\frac{\omega_L}{\epsilon_0\hbar v_g}
\nonumber\\&&\mbox{}
\times\sum_{eg}\big(|d_{eg}^{(1)}|^2+|d_{eg}^{(-1)}|^2\big) |e_\varphi|^2.
\end{eqnarray}
The above expressions show that any dipole-allowed atomic transitions (with $M_e-M_g=0,\pm1$) can contribute to $S_{fxf'x}$
but the $\pi$-type atomic transitions (with $M_e-M_g=0$) cannot contribute to $S_{fyf'y}$.   
The summation rule for the dipole matrix elements of the transitions between the hyperfine manifolds $F$ and $F'$ of the ground state $|nJ\rangle$ and excited state $|n'J'\rangle$, respectively, is \cite{Shore} 
\begin{equation}\label{v13}
\sum_{eg}|d_{eg}^{(q)}|^2=\frac{1}{3}D_{FF'}^2,
\end{equation}
where $q=0,\pm1$ is the label for the spherical tensor components of the electric dipole vector and $D_{FF'}$ is the reduced matrix element of the electric dipole operator of the atom
in the $F$ basis. The expression for $D_{FF'}$ is given by
\begin{equation}\label{v14}
D_{FF'}^2=(2F+1)(2F'+1)
\bigg\{\begin{array}{ccc}
F  &1 &F' \\
J' &I &J
\end{array}\bigg\}^2\langle J\|\mathbf{d}\|J'\rangle^2.
\end{equation}
Here,  $\langle J\|\mathbf{d}\|J'\rangle$ is the reduced matrix element of the electric dipole operator of the atom
in the $J$ basis. With the help of the summation rule (\ref{v13}), we can rewrite Eqs.~(\ref{v12}) as
\begin{eqnarray}\label{v15}
S_{fxf'x}&=&\frac{f}{(\gamma-2i\delta)(2F+1)}\frac{2\omega_LD_{FF'}^2}{3\epsilon_0\hbar v_g}
\nonumber\\&&\mbox{}
\times\big(|e_r|^2+ff'|e_z|^2 \big),
\nonumber\\
S_{fyf'y}&=&\frac{f}{(\gamma-2i\delta)(2F+1)}\frac{2\omega_LD_{FF'}^2}{3\epsilon_0\hbar v_g}|e_\varphi|^2.
\end{eqnarray}


\begin{thebibliography}{99}

\bibitem{Mazur's Nature}  L. Tong, R. R. Gattass, J. B. Ashcom, S. He, J. Lou, M. Shen, I. Maxwell, and  E. Mazur, Nature (London) \textbf{426}, 816 (2003).

\bibitem{Birks} T. A. Birks, W. J. Wadsworth, and P. St. J. Russell,  Opt. Lett. \textbf{25}, 1415 (2000); S. G. Leon-Saval, T. A. Birks, W. J. Wadsworth, P. St. J. Russell, and  M. W. Mason,
in \textit{Conference on Lasers and Electro-Optics (CLEO)},
Technical Digest, Postconference Edition (Optical Society of America, Washington, D.C., 2004),
paper CPDA6. 

\bibitem{taper} J. C. Knight, G. Cheung, F. Jacques, and T. A. Birks, Opt. Lett. \textbf{22}, 1129 (1997); M. Cai and K. Vahala, \textit{ibid.} \textbf{26}, 884 (2001).

\bibitem{Bures99} J. Bures and R. Ghosh, J. Opt. Soc. Am. A \textbf{16}, 1992 (1999).

\bibitem{Tong04} L. Tong, J. Lou, and E. Mazur, Opt. Express \textbf{12}, 1025 (2004).

\bibitem{fibermode} Fam Le Kien, J. Q. Liang, K. Hakuta, and V. I. Balykin, 
Opt. Commun. \textbf{242}, 445 (2004).

\bibitem{Morrissey13} M. J. Morrissey, K. Deasy, M. Frawley, R. Kumar, E. Prel, L. Russell, V. G. Truong, and S. N. Chormaic, Sensors \textbf{13}, 10449 (2013).

\bibitem{fiber trap}
V. I. Balykin, K. Hakuta, Fam Le Kien, J. Q. Liang, and  M. Morinaga,  Phys. Rev. A \textbf{70}, 011401(R) (2004); Fam Le Kien, V. I. Balykin, and K. Hakuta, \textit{ibid.} \textbf{70}, 063403 (2004).

\bibitem{Vetsch10} E. Vetsch, D. Reitz, G. Sagu\'{e}, R. Schmidt, S. T. Dawkins, and A. Rauschenbeutel, Phys. Rev. Lett. \textbf{104}, 203603 (2010).

\bibitem{Goban12} A. Goban, K. S. Choi, D. J. Alton, D. Ding, C. Lacro\^{u}te, M. Pototschnig, T. Thiele, N. P. Stern, and H. J. Kimble, Phys. Rev. Lett. \textbf{109}, 033603 (2012).

\bibitem{Domokos02} P. Domokos, P. Horak, and H. Ritsch, Phys. Rev. A \textbf{65}, 033832 (2002).

\bibitem{absorption} Fam Le Kien, V. I. Balykin, and K. Hakuta, Phys. Rev. A \textbf{73}, 013819 (2006).

\bibitem{Fam14} Fam Le Kien and A. Rauschenbeutel, Phys. Rev. A \textbf{90}, 023805 (2014).

\bibitem{Nayak07} K. P. Nayak, P. N. Melentiev, M. Morinaga, Fam Le Kien, V. I. Balykin, and K. Hakuta, Opt. Express \textbf{15}, 5431 (2007). 

\bibitem{Nayak09} K. P. Nayak, Fam Le Kien, M. Morinaga, and K. Hakuta, Phys. Rev. A \textbf{79}, 021801(R) (2009). 

\bibitem{Dawkins11} S. T. Dawkins, R. Mitsch, D. Reitz, E. Vetsch, and A. Rauschenbeutel, 
Phys. Rev. Lett. \textbf{107}, 243601 (2011). 

\bibitem{Reitz13} D. Reitz, C. Sayrin, R. Mitsch, P. Schneeweiss, and A. Rauschenbeutel, Phys. Rev. Lett. \textbf{110}, 243603 (2013).

\bibitem{Reitz14} D. Reitz, C. Sayrin, B. Albrecht, I. Mazets, R. Mitsch, P. Schneeweiss, and A. Rauschenbeutel, Phys. Rev. A \textbf{89}, 031804(R) (2014). 

\bibitem{Mitsch14a} R. Mitsch, C. Sayrin, B. Albrecht, P. Schneeweiss, and A. Rauschenbeutel,  Phys. Rev. A \textbf{89}, 063829 (2014).

\bibitem{Mitsch14b} R. Mitsch, C. Sayrin, B. Albrecht, P. Schneeweiss, and A. Rauschenbeutel, arXiv:1406.0896. 

\bibitem{Russell13} L. Russell, R. Kumar, V. B. Tiwari, and S. N. Chormaic, Opt. Commun. \textbf{309}, 313 (2013). 

\bibitem{Stiebeiner09} A. Stiebeiner, O. Rehband, R. Garcia-Fernandez, and A. Rauschenbeutel, 
Opt. Express \textbf{17}, 21704 (2009).  

\bibitem{Yalla12} R. Yalla, Fam Le Kien, M. Morinaga,  and K. Hakuta, Phys. Rev. Lett. \textbf{109}, 063602 (2012).

\bibitem{Schroder12} T. Schr\"{o}der, M. Fujiwara, T. Noda, H.-Q. Zhao, O. Benson, and S. Takeuchi, Opt. Express \textbf{20}, 10490 (2012).

\bibitem{Liebermeister13} L. Liebermeister, F. Petersen, A. V. M\"{u}nchow, D. Burchardt, J. Hermelbracht, T. Tashima, A. W. Schell, O. Benson, T. Meinhardt, A. Krueger, A. Stiebeiner, A. Rauschenbeutel, H. Weinfurter, and M. Weber,
Appl. Phys. Lett. \textbf{104}, 031101 (2014).

\bibitem{Brambilla07} G. Brambilla, G. S. Murugan, J. S. Wilkinson, and D. J. Richardson, Opt. Lett. \textbf{32}, 3041 (2007).

\bibitem{Skelton12} S. E. Skelton, M. Sergides, R. Patel, E. Karczewska, O. M. Marag\'{o}, and P. H. Jones, 
J. Quant. Spectrosc. Radiat. Transfer \textbf{113}, 2512 (2012).

\bibitem{Fam13} Fam Le Kien and A. Rauschenbeutel, Phys. Rev. A \textbf{88}, 063845 (2013).

\bibitem{Chormaic14} M. C. Frawley, I. Gusachenko, V. G. Truong, M. Sergides, and S. Nic Chormaic,  Opt. Express \textbf{22}, 16322 (2014). 

\bibitem{Schlosser} N. Schlosser, G. Reymond, I. Protsenko, and P. Grangier,
Nature (London) \textbf{411}, 1024 (2001).

\bibitem{Kuhr} S. Kuhr, W. Alt, D. Schrader, M. M\"{u}ller, V. Gomer, and D. Meschede, 
Science \textbf{293}, 278 (2001). 

\bibitem{Sackett} C. A. Sackett, D. Kielpinski, B. E. King, C. Langer, V. Meyer, C. J. Myatt, M. Rowe, Q. A. Turchette, W. M. Itano, D. J. Wineland, and C. Monroe,
Nature (London)  \textbf{404}, 256 (2000).

\bibitem{Fam09} Fam Le Kien and K. Hakuta, Phys. Rev. A \textbf{79}, 013818 (2009).

\bibitem{Deutsch95} I. H. Deutsch, R. J. C. Spreeuw, S. L. Rolston, and W. D. Phillips, Phys. Rev. A \textbf{52}, 1394 (1995).

\bibitem{Birkl95} G. Birkl, M. Gatzke, I. H. Deutsch, S. L. Rolston, and W. D. Phillips, Phys. Rev. Lett. \textbf{75}, 2823 (1995). 

\bibitem{Henkel03} G. Boedecker and C. Henkel, Opt. Express \textbf{11}, 1590 (2003).

\bibitem{Artoni05} M. Artoni, G. La Rocca, and F. Bassani, Phys. Rev. E \textbf{72}, 046604 (2005). 

\bibitem{Schilke11} A. Schilke, C. Zimmermann, P. W. Courteille, and W. Guerin,  Phys. Rev. Lett. \textbf{106}, 223903 (2011).

\bibitem{Chang07} D. E. Chang, A. S. S{\o}rensen, E. A. Demler, and M. D. Lukin, Nature Phys. \textbf{3}, 807 (2007).

\bibitem{Chang12} D. E. Chang, L. Jiang, A. V. Gorshkov, and H. J. Kimble, New J. Phys. \textbf{14}, 063003 (2012).

\bibitem{Chang11} Y. Chang, Z. R. Gong, and C. P. Sun,  Phys. Rev. A \textbf{83}, 013825 (2011).

\bibitem{Petrosyan07} D. Petrosyan,  Phys. Rev. A \textbf{76}, 053823 (2007).

\bibitem{Schilke12} A. Schilke, C. Zimmermann, and W. Guerin,  Phys. Rev. A \textbf{86}, 023809 (2012).

\bibitem{Ritsch14a} S. Ostermann, M. Sonnleitner, and H. Ritsch,  New J. Phys. \textbf{16}, 043017 (2014).

\bibitem{Ritsch14b} D. Holzmann, M. Sonnleitner, and H. Ritsch,  arXiv:1409.5307.

\bibitem{Fam07} Fam Le Kien and K. Hakuta, Phys. Rev. A \textbf{75}, 013423 (2007). 

\bibitem{coolingbook} H. J. Metcalf and P. van der Straten, \textit{Laser Cooling and Trapping} (Springer, New York, 1999).

\bibitem{Shore} See, for example, B. W. Shore, \textit{The Theory of Coherent Atomic Excitation} 
(Wiley, New York, 1990).

\bibitem{cesium decay} Fam Le Kien, S. Dutta Gupta, V. I. Balykin, and K. Hakuta, 
Phys. Rev. A \textbf{72}, 032509 (2005).

\bibitem{fiber books} See, for example, 
D. Marcuse, \textit{Light Transmission Optics} 
(Krieger, Malabar, FL,  1989); K. Okamoto, \textit{Fundamentals of Optical Waveguides} (Academic Press, New York, 2006); A. W. Snyder and J. D. Love, \textit{Optical Waveguide Theory} (Chapman and Hall, New York, 1983).

\bibitem{Loudon} R. Loudon, \textit{The Quantum Theory of Light}  (Oxford University Press, New York, 2000).

\bibitem{continuous} C. M. Caves and D. D. Crouch, J. Opt. Soc. Am. B \textbf{4}, 1535 (1987);
K. J. Blow, R. Loudon, S. J. D. Phoenix, and T. J. Shepherd, Phys. Rev. A \textbf{42}, 4102 (1990).

\bibitem{fibercorr} Fam Le Kien and K. Hakuta, Phys. Rev. A \textbf{77}, 033826 (2008).

\bibitem{path ordering} L. D. Landau and  L. M. Lifshitz, \textit{Quantum Mechanics: Non-Relativistic Theory} (Course of Theoretical Physics, Vol. 3) (Butterworth-Heinemann, New York, 1991).

\bibitem{Chang14} D. E. Chang, private communication.

\bibitem{Tromborg} T. S{\o}ndergaard and B. Tromborg, Phys. Rev. A \textbf{64}, 033812 (2001).

\end{thebibliography}
\end{document}